\documentclass[preprint,11pt]{elsarticle}
\usepackage{amsmath}
\usepackage{mathtools}
\usepackage{url}
\usepackage{hyperref}

\date{\normalsize\today}

\begin{document}

\begin{frontmatter}

\title{Analysing pandemics in phase-space}

\author[First]{Merlo Olivier \corref{cor1}}

\ead{mero@zhaw.ch}

\cortext[cor1]{Corresponding author}

\address[First]{ZHAW life Science and Facility Management, Gruentalstrasse 14, Postfach , 8820 W\"adenswil, Switzerland}

\begin{abstract}
Based on the SIRD-model a new model including time-delay is proposed for a
description of the outbreak of the novel coronavirus Sars-CoV-2 pandemic. All data were analysed by representing all quantities as a function of the susceptible population, as opposed to the usual dependence on time. 
The total number of deaths could be predicted for the first, second and third wave of the pandemic in Germany with an accuracy of about 10\%, shortly after the maximum of infectious people was reached.
By using the presentation in phase space, it could be shown that a classical SEIRD- and SIRD-model with constant parameters will not be able to describe the first wave of the pandemic accurately.

\end{abstract}


\end{frontmatter}

\section{Introduction}

Since the outbreak of the pandemic in January 2020 with the new virus SARS-CoV-2 in China, a lot of scientists have analysed the obtained data. The data have been evaluated using many methods and 
models. The outbreak was analysed with well known models, like the SIR-model \cite{SIRA}, the SEIRD-model\cite{SEIRDA,SEIRDB}, time-delayed SIR-like models \cite{ebra2020,Devi2021} or other models \cite{MODELA,SEIRAI,MODELC}. 
 Now the virus is spread over the whole world and in some countries several waves of infections has been reported.\\

The model presented is a minor modification of the well-known SIRD model, and as for the SIR model, analytical solutions can be obtained.
Our goal is to determine and predict important quantities of a pandemic. However, we are aware that our simple model cannot describe more complicated situations, such as the  occurrence of a new wave. Additionally, we have to mention that the model 
presented will not give an accurate time dependence of the pandemic, but this will not bother us, because we are interested in long time predictions.\\

The paper is organized as follows. In Section \ref{Model}, we will discuss the advantage of the representation in phase space. Afterwards, we propose a simple model and the analytical solutions of this model. 
The analysing process is shown in Section \ref{analyze} by analysing the data from Germany by John Hopkins University using the results obtained in Section \ref{Model}. In addition, we have analysed the data from Austria, Italy and Switzerland. The results show that simple SEIRD and 
SIRD models with constant parameters cannot adequately describe the pandemic in Germany. In addition, we calculate some key parameters, for example the mortality rate in symptomatic cases or the time from the onset of 
symptoms to death or until recovery. Finally, in this section, we estimate the impact of the lockdown in Germany on the parameters of our model and 
the basic reproduction number $R_0$.\\

In Section \ref{prediction} we show how some key quantities of the pandemic can be predicted and from what moment of the pandemic a good estimate of these quantities can be obtained.

\section{Model and representation}

\subsection{Mathematical Model}
\label{Model}
The SIR-model was introduced by Kermack and Kendrick \cite{SIR} and it divides the population in different groups (susceptible, infectious and recovered) and specifies how the transfer between the 
different groups takes place. In order to include the deceased people in the model, the SIRD-model was introduced, which is an
extension of the SIR-model. The SIRD-model is given 
by the equations \ref{eq:SIRD}, where $S$ stands for `Susceptible`, $I$ for `Infectious`, $R$ for `Recovered` and $D$ for deceased people. In this model, $\beta$ describes the virus transition rate
from person to person and $\gamma_D$ or $\gamma_R$ the transition rate from infectious to deceased or recovered people.\\
 
\begin{align}
\begin{split}
 \dot{S} &=-\beta\cdot S\cdot I\\
 \dot{I} &=\beta\cdot S\cdot I-\gamma_D I-\gamma_R I\\
 \dot{D} &=\gamma_D I\\
 \dot{R} &=\gamma_R I\\
 \end{split}
 \label{eq:SIRD}
\end{align}

If we introduce for all groups the percentage of the whole population $N_0$ then the equations \ref{eq:SIRD} changes slightly to the equations \ref{eq:SIRDp}, where the lower case $p$ indicates the 
reference to the whole population and $\beta_p$ stands for $\beta\cdot N_0$.

\begin{align}
\begin{split}
 \dot{S}_p &=-\beta_p\cdot S_p\cdot I_p\\
 \dot{I}_p &=\beta_p\cdot S_p\cdot I_p-\gamma_D I_p-\gamma_R I_p\\
 \dot{D}_p &=\gamma_D I_p\\
 \dot{R}_p &=\gamma_R I_p\\
\end{split} 
\label{eq:SIRDp}
 \end{align}
 
It is known that any transition between states takes time, for example the incubation time or the time from the onset of illness to death. To include an incubation time into the model, the SEIRD-model (see equation 
\ref{eq:SEIR}) was introduced, where $E$ stands for the `exposed` people and the parameter $\mu$ is the transition rate between the exposed and the infectious people.\\
 
\begin{align}
\begin{split}
 \dot{S} &=-\beta\cdot S\cdot I\\
 \dot{E} &=\beta \cdot S\cdot I-\mu E\\
 \dot{I} &=\mu E-\gamma_D I-\gamma_R I\\
 \dot{D} &=\gamma_D I\\
 \dot{R} &=\gamma_R I\\
 \end{split}
 \label{eq:SEIR}
\end{align}

We will show in the Section \ref{analyze} that the data from SARS-CoV-2 from Germany suggests that a simple SEIRD model is not a good choice to describe the pandemic. Thus, we want to include the transition time by introducing 
the percentage of infectious at time $t-\tau_i$ for the 
transitions from the state 'infectious' to the states 'dead' or 'recovered', see \cite{ebra2020,Devi2021}. The new model with the 2 delay-times $\tau_D$ and $\tau_R$ is shown in equation \ref{eq:SIRDdel}. From these equations it is visible that 
the sum of the susceptible, the infectious, the deceased and the recovered people is constant and therefore a first integral.

\begin{align}
\begin{split}
 \dot{S}_p &=-\beta_p \cdot S_p\cdot I_p\\
 \dot{I}_p &=\beta_p \cdot S_p\cdot I_p-\gamma_D I_p(t-\tau_D)-\gamma_R I_p(t-\tau_R)\\
 \dot{D}_p &=\gamma_D I_p(t-\tau_D)\\
 \dot{R}_p &=\gamma_R I_p(t-\tau_R)\\
 \end{split}
 \label{eq:SIRDdel}
\end{align}

While integrating this model, we have to be aware that the values of the susceptible $S_p$, the infectious $I_p$, the deceased $D_p$ or the recovered population $R_p$ do not have to be positive for all times. The reason for this 
is that a time-delay has 
been introduced, and thus a negative value of $\dot{I}_p$ can be obtained in equation \ref{eq:SIRDdel} even if the value of $I_p(t)$ is already 0 or negative. To be correct, we have to stop the integration the moment the infected population is 0, because 
this marks the end of the pandemic.\\
To be able to obtain an analytical solution, a Taylor expansion of the 
equations \ref{eq:SIRDdel} around $\tau_D=0$ and $\tau_R=0$ was made. The obtained differential equations have to be used carefully, because to be valid the time lag has to be sufficiently small and higher order 
derivatives not too large, see \cite{Maz1974}. The Taylor expansion led to the model in equation 
\ref{eq:SIRDdel2} where $\delta$ is given by $(1-\gamma_D \tau_D-\gamma_R \tau_R)$. The factor $\delta$ is a consequence of the substitution of the function $I_p(t-\tau_D)$ and $I_p(t-\tau_R)$ by their
first-order Taylor series.\\

\begin{align} 
\begin{split}
\dot{S}_p &=-\beta_p S_p\cdot I_p\\
 \dot{I}_p &=\frac{1}{\delta}
\left(\beta_p S_p\cdot I_p-\gamma_D I_p-\gamma_R I_p\right)\\
 \dot{D}_p &=\frac{\gamma_D I_p}{\delta}\left(1-\gamma_R(\tau_R-\tau_D)-\beta_p \tau_D S_p\right)\\
 \dot{R}_p &=\frac{\gamma_R I_p}{\delta}\left(1-\gamma_D(\tau_D-\tau_R)-\beta_p \tau_R S_p\right)\\
\end{split}
 \label{eq:SIRDdel2}
 \end{align}

This model describes a SIRD-model with transition times from the state infectious to the states deceased respectively recovered. It is not obvious that the sum of all states in this model is 
still a first integral.\\

We have calculated the basic reproduction number $R_0$ using the method of Driesche et al \cite{Driesche2002} and it is given by  $\protect\mathmbox{R_0 = \protect \beta_p\cdot S_p/(\gamma_R+\gamma_D)
}$, which is exactly the same as in the SIRD-model.\\
We were able to find analytical solutions of this system of ordinary differential equations \ref{eq:SIRDdel2} with the same method as Harko et al \cite{SIRexact} used to solve the SIR-model. The solutions for $I_p$, $D_p$ and $R_p$ in dependence of 
the `susceptible` population $S_p$, where $I_{p_0}$, $D_{p_0}$ respectively $R_{p_0}$ are the initial conditions at time 0 are then given by the equations \ref{eq:SIRDdel2exact} if $I_{p_0}$ is not zero. To simplify 
the equations, the new variable $\protect\mathmbox{\lambda=\protect (\gamma_D+\gamma_R)/\beta_p}$ was introduced which is alternativly given by $\protect\mathmbox{\lambda=\protect S_p/R_0}$.\\

\begin{align}
\begin{split}
 I_p &=\frac{(S_{p_0}-S_p)}{\delta}+I_{p_0}+\frac{\lambda}{\delta}\ln\left(\frac{S_p}{S_{p_0}}\right)\\
 D_p &=\frac{\gamma_D \tau_D (S_p-S_{p_0})}{\delta}+D_{p_0}-\frac{\gamma_D (1+\gamma_R (\tau_D-\tau_R))}{\beta_p\cdot \delta}
 \ln\left(\frac{S_p}{S_{p_0}}\right)\\
 R_p &=\frac{\gamma_R \tau_R (S_p-S_{p_0})}{\delta}+R_{p_0}-\frac{\gamma_R (1+\gamma_D (\tau_R-\tau_D))}{\beta_p\cdot \delta}
 \ln\left(\frac{S_p}{S_{p_0}}\right)\\
\end{split}
 \label{eq:SIRDdel2exact}
 \end{align}

To calculate the relation between time and `susceptible` population we have to integrate $\protect\mathmbox{\protect
1/\dot{S}_{p}
}$ from $S_{p_0}$ to $S_p$, see Harko et al \cite{SIRexact}.

\begin{align}
 t &=t_0+\int\limits_{S_{p_0}}^{S_p}\frac{1}{\dot{S}_{p}} dS_p=t_0+\int\limits_{S_{p_0}}^{S_p}\frac{1}{\beta_p\cdot S_p\cdot I_p(S_p)}dS_p
 \label{eq:SIRDdel2exacttime}
 \end{align}

The poles $S_{p_{pol}}$ of $\protect\mathmbox{\protect
1/\dot{S}_{p}
}$ defines the asymptotic percentage of people $S_{p_{\infty}}$ which will not be infected in the limit for infinite time. A short calculation gives a percentage of 
$\protect\mathmbox{S_{p_{\infty}}=\protect
-\lambda\cdot W_0\left(\exp(-\frac{1}{\lambda}(S_{p_0}+I_{p_0}\delta))\cdot (-S_{p_0})/\lambda\right)
}$ or $S_{p_{\infty}}=0$ with the Lambert function $W_0(x)$. The reason why the poles define this limit is that the time is calculated by $\protect\mathmbox{\int(\protect 1/\dot{S}_{p}
) dS_p}$ and around the pole the Taylor expansion of the denominator is proportional to $(S_p-S_{p_{pol}})^2$ 
respectively  $(S_p-S_{p_{pol}})^1$ and this means that 
the time $t$ will go to infinity, if we approximate a pole\footnote{For $I_{p_0}=0$. It follows from the equation \ref{eq:SIRDdel2exact} for $I_p$ that a pole is given by $S_{p_0}$ and then the asymptotic value of $S_{p}$ is given by $S_{p_{\infty}}=S_{p_0}$.}.

The percentage of people who will die or will recover if the time goes to infinity $D_{p_{\infty}}$ or $R_{p_{\infty}}$ are calculated by inserting $S_{p_{\infty}}$ into the corresponding equations \ref{eq:SIRDdel2exact}.\\

The linearised form of the simplest model with time-delay, see equations \ref{eq:SIRDdel2s}, leads to the same form of solutions of our new model \ref{eq:SIRDdel2exact} with different coefficients before the functions of $S_p$ and $\ln(S_p)$. 
Using the data and methods presented in section \ref{analyze}, it can be shown that the simplest model is not suitable to describe the pandemic, as it leads to some impossible values of the parameters.    

\begin{align}
\begin{split}
 \dot{S}_p &=-\beta_p \cdot S_p\cdot I_p(t-\tau)\\
 \dot{I}_p &=\beta_p \cdot S_p\cdot I(t-\tau)_p-\gamma_D I_p-\gamma_R I_p\\
 \dot{D}_p &=\gamma_D I_p\\
 \dot{R}_p &=\gamma_R I_p\\
 \end{split}
 \label{eq:SIRDdel2s}
\end{align}

\subsection{Phase space}

In these kinds of systems, only two of the four quantities, namely the susceptible and the infectious people, are important to describe the dynamics of the pandemic. 
This is because the susceptible people are those who can be infected and the infectious are those who can infect others. If we want to describe the quantities of the system in a kind of phase space in dependence of one of 
the quantities. 
Then the susceptible people is the better choice of the two because this 
quantity is usually strictly decreasing during a wave, in contrast to the number of infectious people.\\

The other two quantities, the recovered and the deceased people will describe the dynamic in simple models like the SIR model, too, but the recovered people will not work in a bit more complicated cases. 

In the following we consider the corresponding variables, such as infected, recovered and deceased, always depending on the
susceptible people and not as a function of time. We want to exploit a property of this representation, namely that a time transformation with an adjustment of the system parameters maps a whole family of solutions 
onto the same orbit in this representation. As we will see later, this representation usually makes it much easier
to make comparisons between different models for the same pandemic. In our case, we even have analytical results for the different models considered, making it easier to choose the model that best describes the data.

\section{Application of the mathematical model to the SARS-CoV-2 pandemic in Germany}
\label{analyze}

In this section we will always analyse the data using our model, except in two cases where we analyse the data using some analytical solutions of the SEIRD resp. SIRD model to show that
these two models do not describe the data well. 

\subsection{Source of Data and Data manipulation}

In our model \ref{eq:SIRDdel2} we need the number of susceptible, infectious, recovered and deceased people. All the accessible data does not include the susceptible population and a lot of them do not contain the 
recovered people or are of poor quality. Thus, we decided to apply 
our method to the data from Germany. The data from Switzerland, 
Austria and Italy were investigated, too, but give no new insight. The results from Italy have to be interpreted carefully because the first wave of the pandemic was almost over at the $22^{nd}$ of June 2020
, 
so you have a lot of data from this part of the curve in this representation and therefore the `ending` part  
of the curve is over-represented, see Figure \ref{fig:yvx}.\\

The time-series data of the John-Hopkins-University \cite{JHU} contains the date $t$, the cumulative sum of the infected $I_{\Sigma}$, the cumulative sum of the dead $D$ and the cumulative sum of the recovered people $R$ on a daily basis. 
If we know  the total population $N_0$ then we are able to calculate the percentage of susceptible $\protect\mathmbox{S_p=\protect
(N_0-I_{\Sigma})/N_0
}$, of infectious $\protect\mathmbox{I_p=\protect
(I_{\Sigma}-D-R)/N_0
}$, of dead $\protect\mathmbox{D_p=
\protect D/N_0
}$ and 
of recovered people $\protect\mathmbox{R_p=\protect R/N_0}$ for every day. In our analyses we have used the data up to the date of $22^{nd}$ of June 2020
 for the analysis of the first wave in the Sections \ref{sec:accpop}, \ref{sec:extparam}, \ref{sec:estgor} 
and \ref{sec:predfirst} and the data up to the $8^{th}$ of August 2021 for the analysis of the second and third wave, see Section \ref{sec:2wave}.\\

As mentioned above, we need to know the total number of people $N_0$ in order to analyse the data, but this is fundamentally not possible. It cannot be the entire population $N_p$ of the country because the pandemic
starts locally and thus at least at this point in time of the pandemic the entire population of the country cannot be infected. The time dependence 
of the accessible population cannot be estimated because it depends on the population density of the country, the mobility of the population and  a lot of other quantities and especially where in this population density 
the pandemic begin to spread.
In our analysis, we will set this accessible population $N_0$ to the same value for the entire data set for all times in one analysing process. It is surprising that this approach works very well, even if we are using a 
much too high population at the start of the pandemic. By using a perturbation approach we are able to understand this behaviour \footnote{In all solutions of our model \ref{eq:SIRDdel2exact} the left side and the first 
part on the right hand side of the equation are proportional to $1/N_0$. A perturbation calculation shows that $ \ln\left(S_p/S_{p_0}\right)$ 
 is proportional to $N_0^{-1}$ if we assume that at the beginning of the pandemic $S_p$ is approximately $1-I_p$ and that $I_p$ is much smaller than 1.}.

\subsection{Calculating the accessible population $N_0$}
\label{sec:accpop}

We will use our analytical results \ref{eq:SIRDdel2exact} to estimate the accessible population $N_0$ of Germany. Taking a look at the equations \ref{eq:SIRDdel2exact} we recognize that the dependence of all the quantities $I_p$, $D_p$ and $R_p$ of $S_p$ 
is of the form $y=a+b\cdot x+c\cdot \ln(x)$ with different 
coefficients $a$, $b$ and $c$. To be better memorable we will use the indices $x^0$, $_x$ or $_{lnx}$ instead of the variables $a$, $b$ or $c$ for the different coefficients. Additionally, we have to 
indicate the plot to which we refer to by using the notation $I_p$, $D_p$ and $R_p$. Therefore, $I_p(S_p)$ is given by $I_{p_{x^0}}+I_{p_x}\cdot x+I_{p_{lnx}}\cdot \ln(x)$.\\

\begin{figure}[h]
\protect\centering
\hspace{-0.5cm}\includegraphics[width=4.3cm]{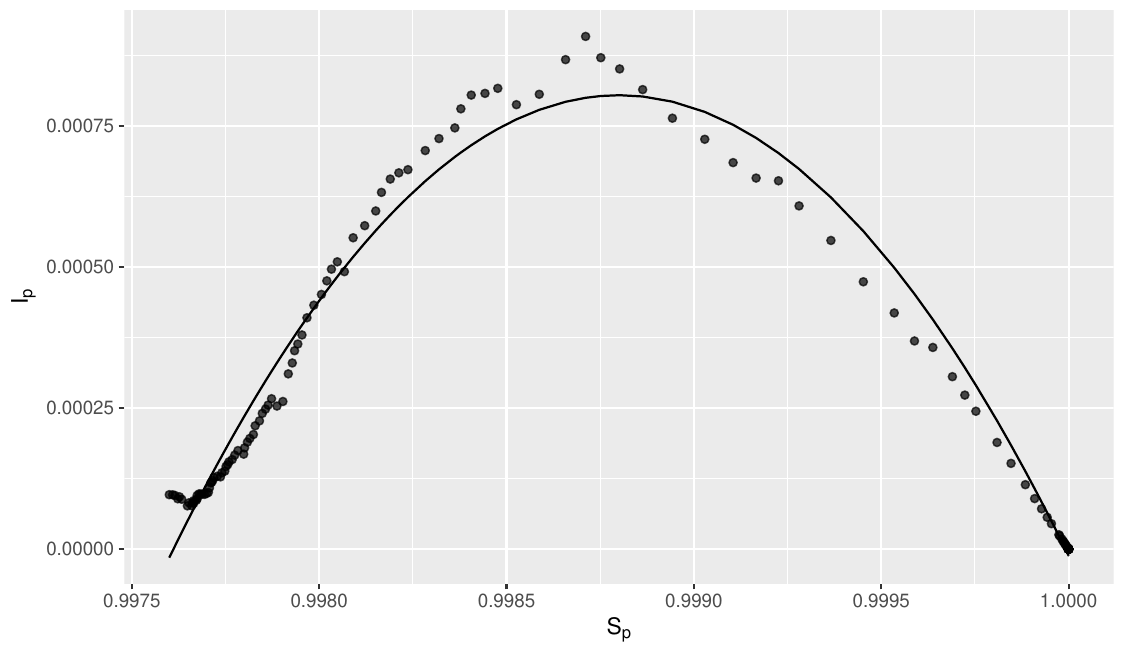}
\includegraphics[width=4.3cm]{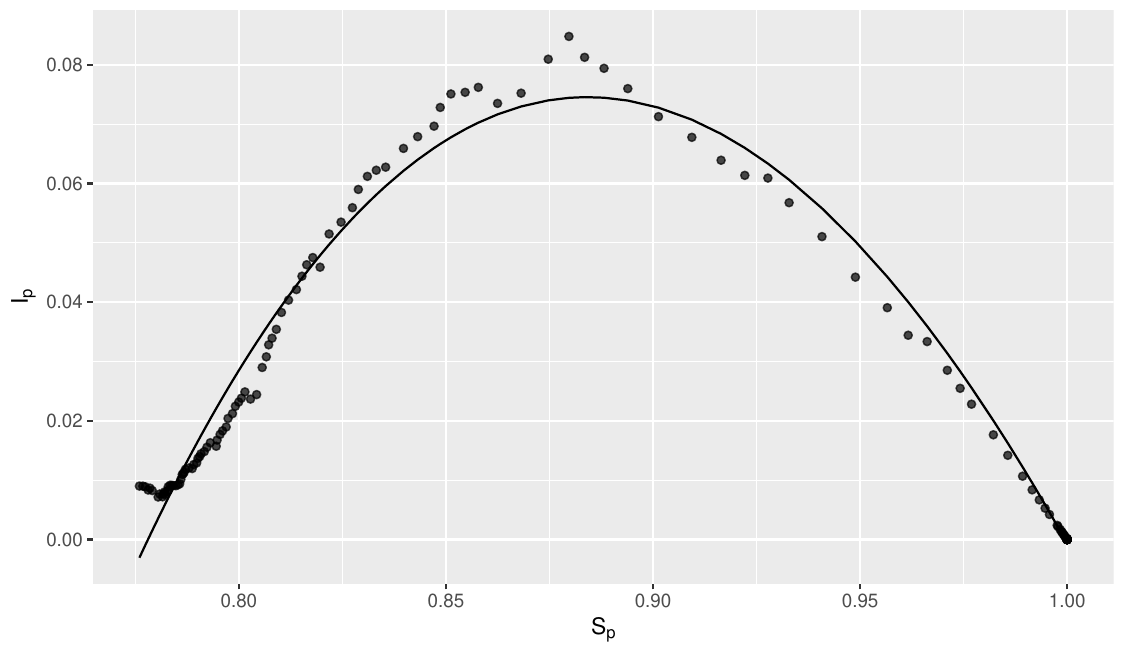}
\includegraphics[width=4.3cm]{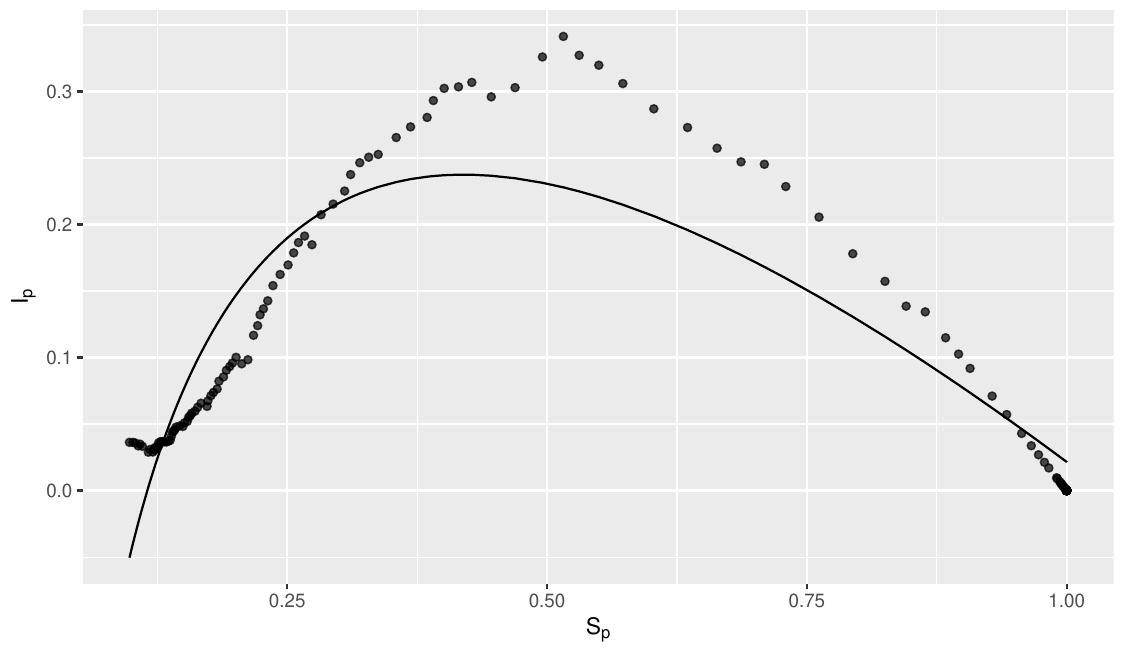}\\
\caption{Here the percentage of the infectious population is plotted against the percentage of the susceptible population using different numbers of the accessible population. On the left, an accessible population of $\protect\mathmbox{N_0=83\cdot 10^{6}}$ was used 
which coincide with the population of Germany. In the middle, the accessible population $\protect\mathmbox{N_0=\protect 859225}$ was taken by maximizing the 
coefficient of determination $\protect\mathmbox{R^2}$. On the right side, the accessible population $\protect\mathmbox{N_0 = \protect 213440 }$ was used to meet the criteria that the adapted function should take the functional 
form of the SIRD-model, which is given by $\protect\mathmbox{I_p =\tilde{a}-S_p+\tilde{c}\ln(S_p)}$ and that
 the coefficient of determination $\protect\mathmbox{R^2}$ is maximized. Then the data is best reproduced by 
$I(p)=1.0215-S_p+0.4196\cdot \ln(S_p)$.}
\label{fig:yvx}
\end{figure}

We will use the dependence of the infectious population $I_p$ from the susceptible population $S_p$ to estimate the accessible population $N_0$. In a first step, we have plotted on the left in Figure \ref{fig:yvx} the infectious versus the susceptible population 
 using the data from the John Hopkins University and an accessible population of $N_0=83\cdot 10^{6}$ which is the whole population of Germany. To clarify it, we have obtained the susceptible population $\protect\mathmbox{S_p=\protect (N_0-I_{\Sigma})/N_0}$ by subtracting 
 the total number of infected $I_{\Sigma}$ from the accessible population $N_0$ and afterwards dividing it by the accessible population. In the middle 
of Figure \ref{fig:yvx} we have chosen the accessible population ($N_0=859225$) maximizing the coefficient of determination  $R^2$\cite{Sachs2015} by fitting the 
function $y=a+b\cdot x+c\cdot \ln(x)$ to the data with the restriction that the number of people are at least the maximal number of confirmed cases and are maximal the 
population of the whole Germany. This specific function was taken because our models predict such a dependence. On the right-hand 
side we have obtained an accessible population of $N_0=
213440$ by maximizing the coefficient of determination in case the coefficient $I_{p_x}$ is $-1$. This is just the theoretical value for the SIRD-model where the infectious population is given by $I_p\left(S_p\right)=\left(S_{p_0}-S_p\right)+I_{p_0}+\lambda\ln\left(S_p/S_{p_0}\right)$with some constant $\lambda$. It is visible on the right-hand side in Figure \ref{fig:yvx} 
 that the SIRD-model with constant parameters will not describe the data adequately.
 
\begin{figure}[h]
\centering
\hspace{-0.5cm}\includegraphics[width=4.3cm]{GermanyN0r2IvsS}
\includegraphics[width=4.3cm]{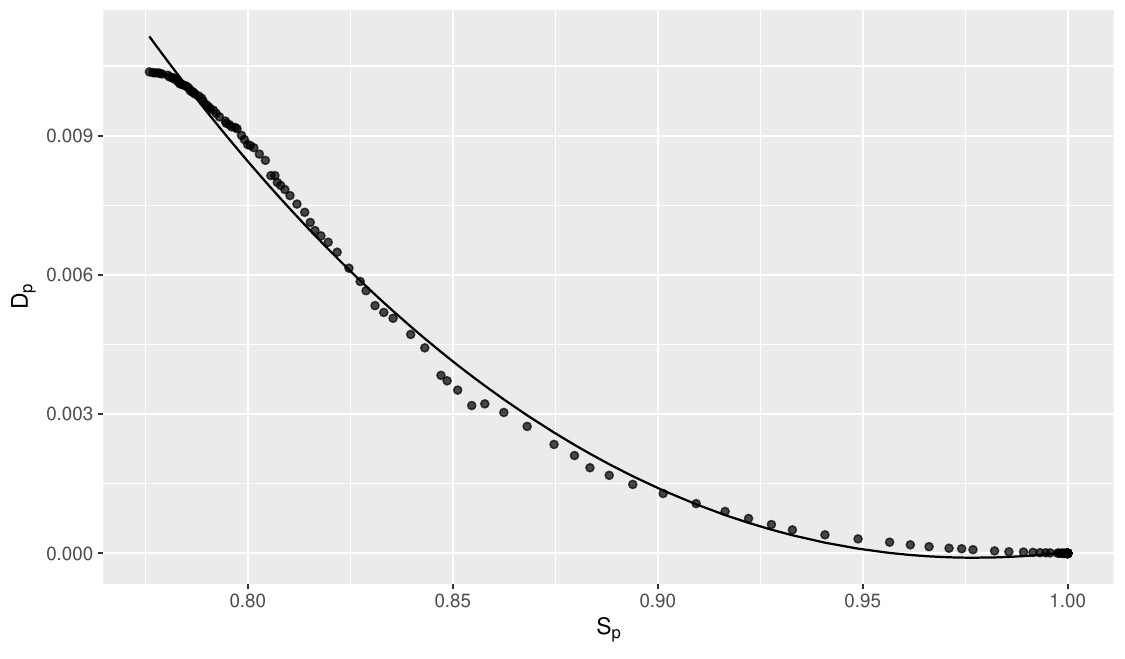}
\includegraphics[width=4.3cm]{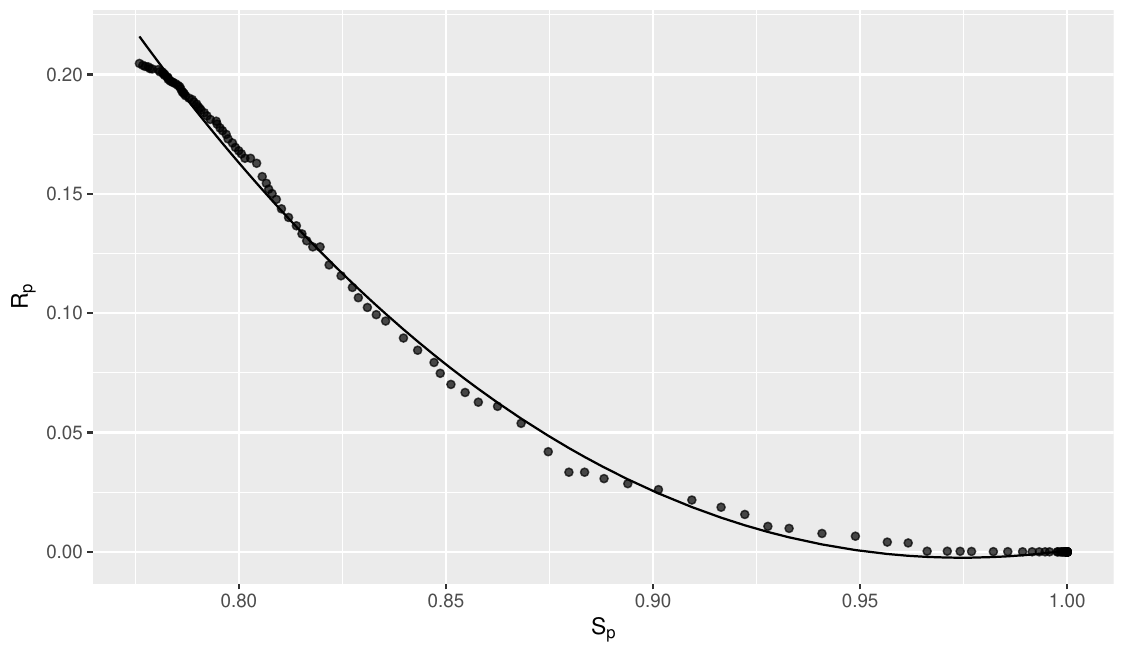}\\
\caption{From left to right, we have plotted with an accessible population of $N_0=\protect 859225 $ the data of Germany and the fitted function for the infected $I_p$, the deceased $D_p$ and the recovered population $R_p$.}
\label{fig:Germany1}
\end{figure}

The fitted curves in the left and in the middle graphs describes the data quite well, but in the spirit of the accessible population we used in Figure \ref{fig:Germany1} for all plots an accessible population of $N_0=859225 $. 
As can be seen in Figure \ref{fig:Germany1}, not only the infectious, but also the deceased and the recovered population of the pandemic of Germany are described very well with an accessible population of $N_0=859225 $.
It is clearly evident that the data points at the end of the pandemic in every of these plots have some discrepancy from the fitted line.

\begin{figure}[h]
\centering
\includegraphics[width=7cm]{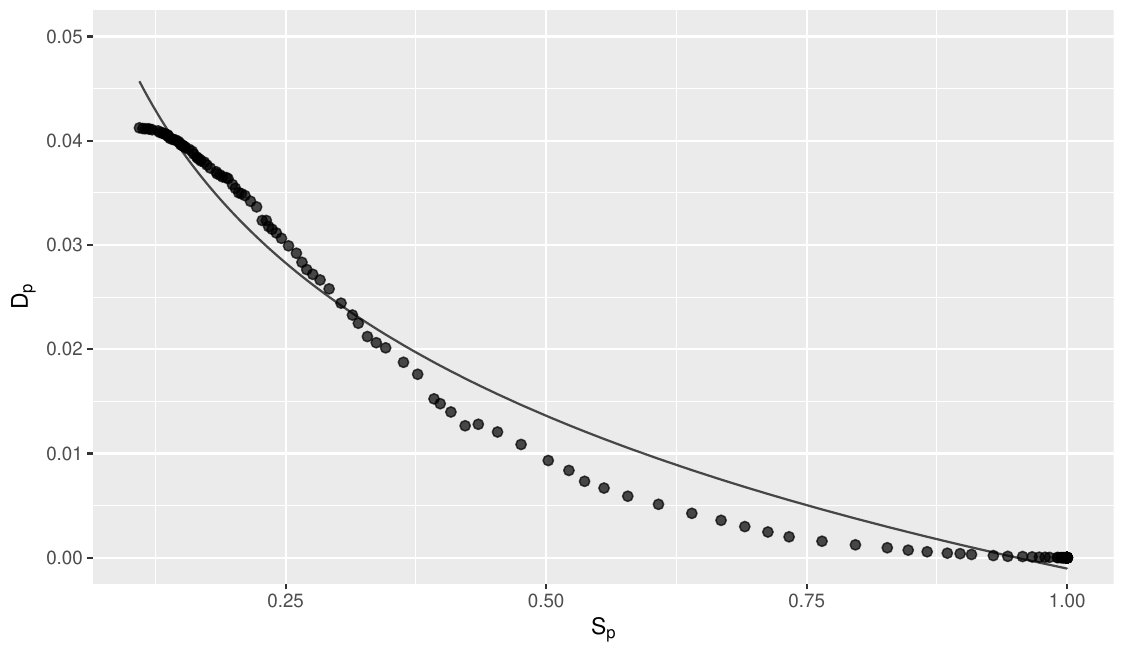}
\caption{We have plotted the data of Germany of the deceased versus the susceptible population using an accessible population of $N_0=\protect 216197 $ which we obtained by maximizing the coefficient of determination for the theoretical function 
$y=\tilde{a}+\tilde{b}\ln(x)$, see text. The data are badly described by the function and therefore a SEIRD-model will not be able to describe the data well.}
\label{fig:SEIRD}
\end{figure}

Before we want to exploit the obtained functions from the fit to estimate the parameters of the model, we want to show that the SEIRD-model will not describe the data well. Since we were not able to find analytical solutions for all quantities for the 
SEIRD-model (see equations \ref{eq:SEIR}), we concentrate on the deceased people for whom the analytical solution is given by $D_p=D_{p_0}-\gamma_{D}/\beta_p \ln\left(S_p/S_{p_0}\right)$. 
Therefore, a function of the form $y=\tilde{a}+\tilde{b}\ln(x)$ should be able to describe the data. As for our model, we determined the accessible population $N_0= 216197 $ by maximizing the coefficient of determination $R^2$ for this function. 
In Figure \ref{fig:SEIRD} it is obvious that the function is not describing the data well and thus the SEIRD-model with constant coefficients is not a good choice to describe this pandemic, too.

\subsection{Estimating the parameters of the model}
\label{sec:extparam}
Now we will exploit the obtained functions from the fits to get the parameters $\beta_p$, $\gamma_D$, $\gamma_R$, $\tau_D$ and $\tau_R$ of our model, see equations \ref{eq:SIRDdel2}.  

\begin{table}[!htbp] \centering 
\scalebox{0.8}{
\begin{tabular}{@{\extracolsep{5pt}} lcccc}
\\[-1.8ex]\hline 
\hline \\[-1.8ex] 
plot & a & b & c & $R^2$\\
\hline \\[-1.8ex] 
$I_p$ vs. $S_p$ & 10.768 & $I_{p_x}=-10.768$ & $I_{p_{lnx}}=9.521$ & 0.978\\
$D_p$ vs. $S_p$ & -0.466 & $D_{p_x}=0.466$ & $D_{p_{lnx}}=-0.456$ & 0.997\\
$R_p$ vs. $S_p$ & -9.301 & $R_{p_x}=9.302$ & $R_{p_{lnx}}=-9.065$ & 0.998\\
\hline \\[-1.8ex]
\end{tabular}
}
\caption{Fitted Functions for Germany.\label{tab:func}}
\end{table}

With the 3 fitted functions (see Table \ref{tab:func}) almost all parameters can be calculated by comparing the coefficients of the fitted functions with those of the analytical solutions \ref{eq:SIRDdel2exact} of our model. 
Due to the first integral $ S_p + I_p + D_p + R_p $ the relations \ref{eq:relations} must be valid for the coefficients of the equations \ref{eq:SIRDdel2exact} respectively the fitted functions of our model and therefore not all parameters of the fitted functions 
are independent.

\begin{align}
\begin{split}
-1 &= I_{p_x}+D_{p_x}+R_{p_x}\\
0  &= I_{p_{lnx}}+D_{p_{lnx}}+R_{p_{lnx}}\\
\end{split}
 \label{eq:relations}
\end{align}

We are using all the coefficients of the fitted functions, leaving out the intercepts, because all these values are in absolute values very close to the coefficients of $x$ in the same fitted function, see table \ref{tab:func}. This leads to 4 independent equations, 
and therefore we are able to get 4 parameters as functions of the last unknown parameter for example $\tau_R$, see equations \ref{eq:param1}. 

\begin{align}
\begin{split}
\gamma_D &=-\frac{R_{p_{x}} (I_{p_{x}} D_{p_{lnx}}+D_{p_{x}} (D_{p_{lnx}}+R_{p_{lnx}}))}{I_{p_{x}}(I_{p_{x}}
   R_{p_{lnx}}+R_{p_{x}} (D_{p_{lnx}}+R_{p_{lnx}}))} \frac{1}{\tau_R}\\
\gamma_R &=-\frac{R_{p_{x}}}{I_{p_{x}}} \frac{1}{\tau_R}\\
\beta_p &= -\frac{R_{p_{x}}
   (I_{p_{x}}+D_{p_{x}}+R_{p_{x}})}{(I_{p_{x}} R_{p_{lnx}}+R_{p_{x}} (D_{p_{lnx}}+R_{p_{lnx}}))}\frac{1}{\tau_R}\\
\tau_D &= \frac{D_{p_{x}}
   (I_{p_{x}} R_{p_{lnx}}+R_{p_{x}} (D_{p_{lnx}}+R_{p_{lnx}}))}{R_{p_{x}} (I_{p_{x}} D_{p_{lnx}}+D_{p_{x}}
   (D_{p_{lnx}}+R_{p_{lnx}}))}\tau_R \\
\end{split}
 \label{eq:param1}
   \end{align}

\begin{figure}[h]
\centering
\hspace{-0.5cm}\includegraphics[width=4.3cm]{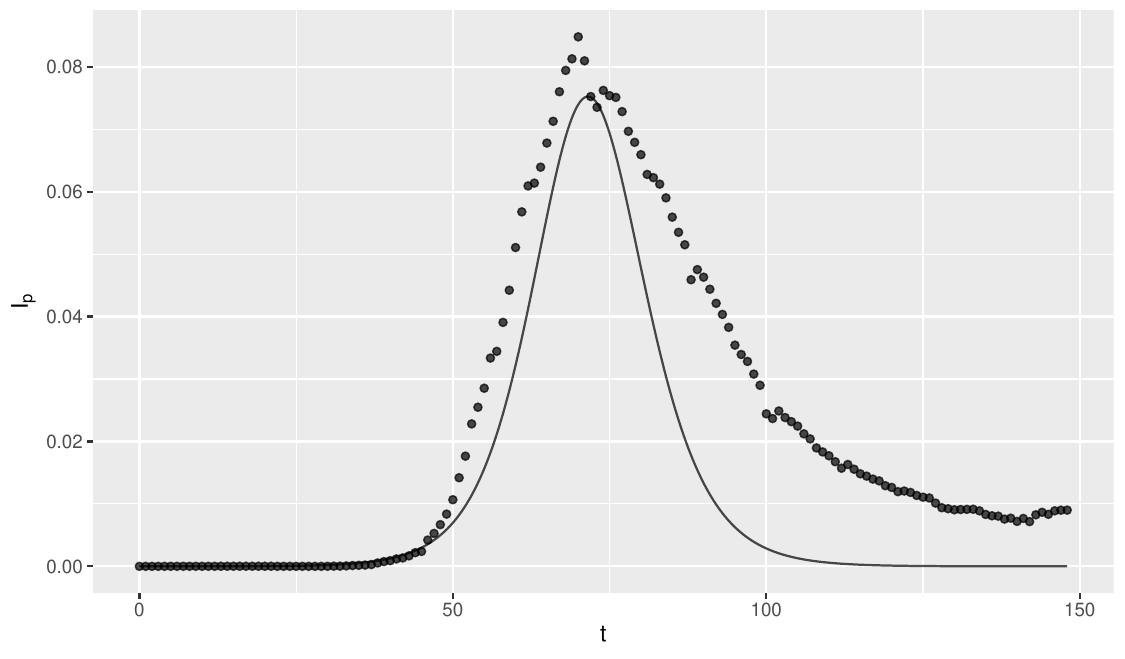}
\includegraphics[width=4.3cm]{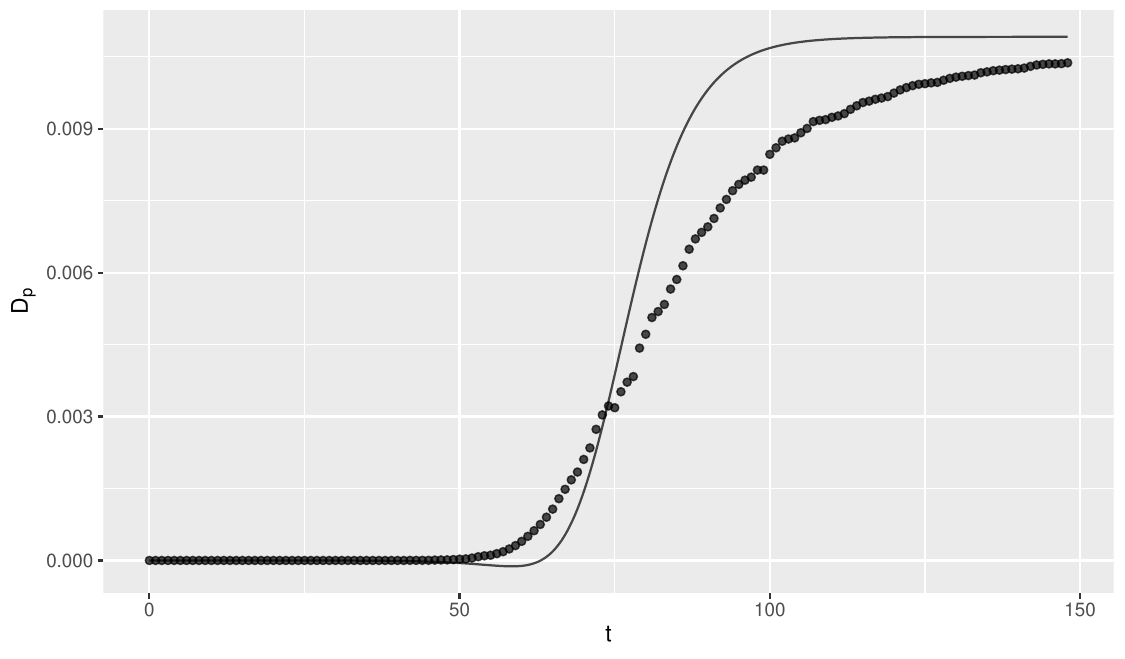}
\includegraphics[width=4.3cm]{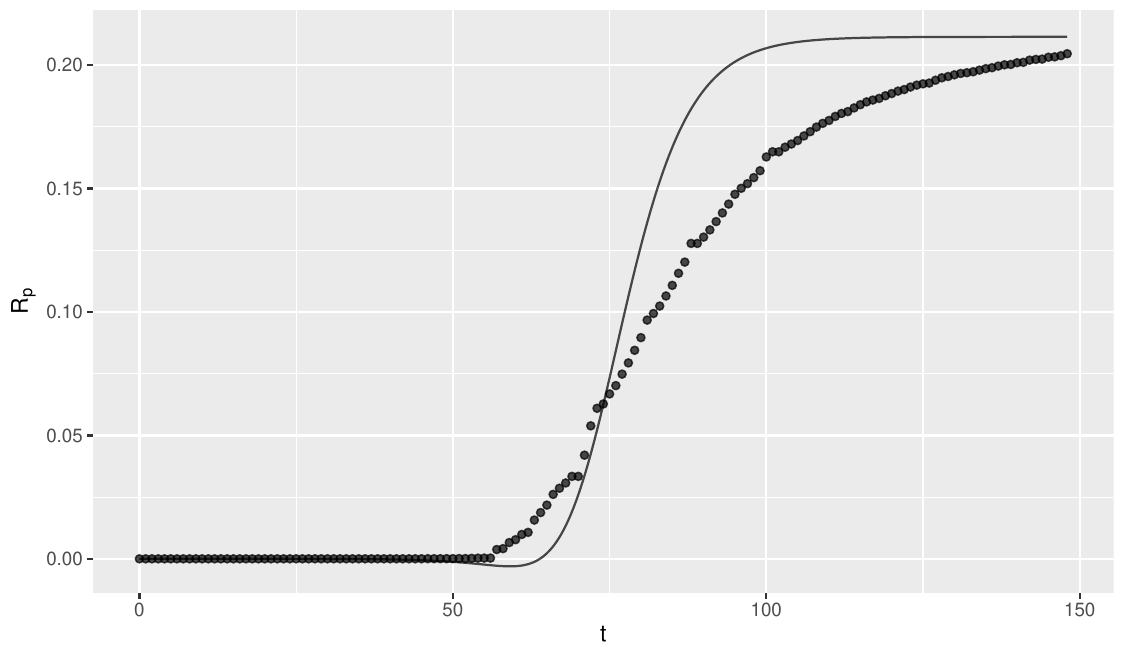}\\
\caption{From left to right, we have plotted the time-dependence of the data of Germany for the infectious $I$, the deceased $D$ and the recovered population $R$ and the solution of the ode with the parameter given in Table \ref{tab:Europe} . }
\label{fig:germanyvont}
\end{figure}

To fix the last parameter 
$\tau_R$ we are solving the ordinary differential equations \ref{eq:SIRDdel2} numerically and fixing $\tau_R$ in the way that the time of the occurrence of the maxima of the infectious population $I_p$ coincide with the one 
obtained from the data. The parameters of our model 
\ref{eq:SIRDdel2exact} estimated 
by this method are shown in Table \ref{tab:Europe} and the plots of the data and the corresponding solutions of the ordinary differential equations are shown in Figure \ref{fig:germanyvont}. It can be seen that the time dependence of the 
epidemic by the model with the parameters found is not well described. Because we are interested in forecasts for 
very long periods of time and not in short-term forecasts, this does not bother us. Since the percentage of deceased resp. recovered people converge towards a value, the long term behaviour will be well predicted.\\

\begin{table}[!htbp] \centering 
 \scalebox{0.75}{
 \hspace{-0.5cm}
\makebox{
\begin{tabular}{@{\extracolsep{5pt}} cccccccccc} 
\\[-1.8ex]\hline 
\hline \\[-1.8ex] 
Country & $N_p$ & $N_0$ & $\beta_p$ & $\gamma_D$ & $\gamma_R$ & $\tau_D$ & $\tau_R$ & $\delta$ & $\lambda$\\ 
\hline \\[-1.8ex] 
Germany & 80200000 & 859225  & 0.140 & $6.09\cdot 10^{-3}$ & 0.118 & 7.1  & 7.3  & 0.093 & 0.884 \\ 
Switzerland & 8400000 & 40378  & 0.247 & $8.27\cdot 10^{-3 }$ & 0.121 & 2.4 & 4.8 & 0.396 & 0.523 \\ 
Austria & 8900000 & 43610  & 0.150 & $4.87\cdot 10^{-3}$ & 0.114 & 5.7 & 7.5 & 0.124 & 0.788 \\ 
Italy & 62400000 & 248890  & 0.185 & $7.98\cdot 10^{-3}$ & 0.045 & -11.9 & 5.9 & 0.831 & 0.284 \\ 
\hline \\[-1.8ex]
\end{tabular} 

}
}

  \caption{The total population $N_{p}$, accessible population $N_{0}$ and the obtained parameters for Germany, Austria, Italy and Switzerland.  \label{tab:Europe} 
} 
\end{table}

With the parameters found in Table \ref{tab:Europe} for Germany and the initial conditions $S_{p_0}=1-1.164 \cdot 10^{-6} $ and $I_{p_0}=1.164 \cdot 10^{-6}$
we can now calculate the asymptotic percentage of people who will never be infected
$S_{p_{\infty}}= 0.778$, and by inserting this obtained value and the obtained parameters in equation \ref{eq:SIRDdel2exact} we get the percentage of recovered $R_{p_{\infty}} = 0.212 $ respectively dead $D_{p_{\infty}}=  0.011 $ for an infinitely long time. 
Additionally, we are able to calculate the maximal percentage of infectious by finding the maxima of $I_{p_{max}}$ in equation \ref{eq:SIRDdel2exact}, which is given by $I_{p_{max}}=I_p\left(1/\lambda\right)=0.075 $.\\
The transition time from infectious state to dead respectively recovered state is given by the sum of the time delay $\tau_D$ or $\tau_R$ and by the time period in the infectious state. It 
is known that the estimated time in the state $E$ in the SEIR-model is given by $\frac{1}{\mu}$ (see equations \ref{eq:SEIR}). In our model, the mean time in the state $I_p$ is then given by 
$\protect\mathmbox{ \protect \delta/(\gamma_D+\gamma_R) }=0.75$. Therefore, we get a total transition time from state $I_p$ to state $D_p$ or $R_p$ of 
$\tau_{D_{tot}}=0.75+ 7.1 \approx 7.9$ or $\tau_{R_{tot}}=0.75+7.3\approx 8.1 $.\\

We are not able to calculate the infection fatality ratio, because the confirmed cases do not include 
the asymptomatic cases, but we can calculate the case fatality ratio (CFR). Knowing the above quantities the CFR is given by $\protect\mathmbox{CFR= \protect D_{p_{\infty}}/(D_{p_{\infty}}+R_{p_{\infty}})}$. In Table \ref{tab:Europe} a summary of the analyses of Germany, Austria, Italy and Switzerland is shown. It 
contains the population, the accessible population and the obtained parameters.\\

\begin{table}[!htbp] \centering 
 \scalebox{0.75}{
 \hspace{-0.5cm}
\makebox{

\begin{tabular}{@{\extracolsep{5pt}} cccccccc} 
\\[-1.8ex]\hline 
\hline \\[-1.8ex] 
Country & $I_{P_{max}}$ & $D_{P_{\infty}}$ & $R_{P_{\infty}}$ & CFR & $R_0$ & $\tau_{D_{tot}}$ & $\tau_{R_{tot}}$ \\ 
\hline \\[-1.8ex] 
Germany & 0.075  & 0.011  & 0.212  & 0.0491  & 1.13  & 7.9  & 8.1  \\ 
Switzerland & 0.345 & 0.049 & 0.727 & 0.0634 & 1.91 & 5.5 & 7.9\\ 
Austria & 0.188 & 0.016 & 0.384 & 0.0407 & 1.27 & 6.7 & 8.5 \\ 
Italy & 0.431 & 0.145 & 0.823 & 0.15 & 3.52 & 3.9 & 21.7 \\ 
\hline \\[-1.8ex] 
\end{tabular} 

}
}
  \caption{In the first few columns, the maximal percentage of infectious $I_{P_{max}}$ of the population, the percentage which will die $D_{P_{\infty}}$  and the percentage which will recover $R_{P_{\infty}}$ of the
  population is shown. Then the CFR, the basic reproduction number with assuming $S_p$ equal 1 and the time from onset of symptons to death $\tau_{D_{tot}}$ or recovery $\tau_{R_{tot}}$ is visible. All quantities were calculated for Germany, 
  Switzerland, Austria and Italy.} 
  \label{tab:Europequant}
\end{table}

A closer look at the Table \ref{tab:Europe} shows that  the $\beta_p$ values of Germany and Austria are close, and the values of Italy and Switzerland are a little bit higher. The $\gamma_D$ values of all countries are 
roughly the same. The same behaviour can be seen with the $\gamma_R $ values, except for Italy, which has a significantly smaller $\gamma_R $ value. This leads to a significantly higher mortality rate.  
In the Table \ref{tab:Europequant} some additional quantities are listed, and again it is clearly visible that the 
countries Germany, Switzerland and Austria are comparable with each other and that Italy has a special position here, too.\\

During a pandemic, usually we are more interested in absolute values or the percentage of the whole population $N_p$ of the country than the percentage of the accessible population $N_0$. Here it would be completely 
unrealistic to multiply the 
obtained percentages by the total population of the country in order to get an estimate of the absolute quantity, because our data is not of this form. We have calculated the estimate of the absolute number of deaths by 
multiplying the corresponding percentage by the accessible population, for example for the number of death persons $D=N_0\cdot D_{p_{\infty}}$ 
and the maximal number of infectious people $I_{max}=I_{p_{max}}\cdot N_0$. The obtained values and the corresponding values calculated from the data are shown in Table \ref{tab:Europequantabs}.

\begin{table}[!htbp] \centering 
 \scalebox{0.75}{
\makebox{

\begin{tabular}{@{\extracolsep{5pt}} ccccc} 
\\[-1.8ex]\hline 
\hline \\[-1.8ex] 
Country&$I_{max}$&$I_{max_{data}}$&$D_{\infty}$&$D_{\infty_{data}}$\\
\hline \\[-1.8ex] 
Germany & 64075 & 72864 & 9403 & 8914 \\ 
Switzerland & 13913 & 14349 & 1987 & 1956 \\ 
Austria & 8212 & 9334 & 711 & 693 \\ 
Italy & 107147 & 108257 & 36070 & 34675 \\ 
\hline \\[-1.8ex] 
\end{tabular} 

}
}
  \caption{The calculated maximal number of infectious people by the model $I_{max}$, using the data $I_{max_{data}}$ and the estimated Number of deaths $D_{\infty}$ and the number of deaths at the last date of the data.} 
  \label{tab:Europequantabs}

\end{table}

It is clear that both sizes match well, since the first wave of the pandemic in Germany is about to end. In the next section, we will see that these quantities start to agree well shortly after the infectious population has exceeded their maximum. 
But first, we will investigate the influence of the governmental measures on the parameters of the model.\

\subsection{Estimating governmental measures}
\label{sec:estgor}
\begin{figure}[h]
\centering
\includegraphics[width=10cm]{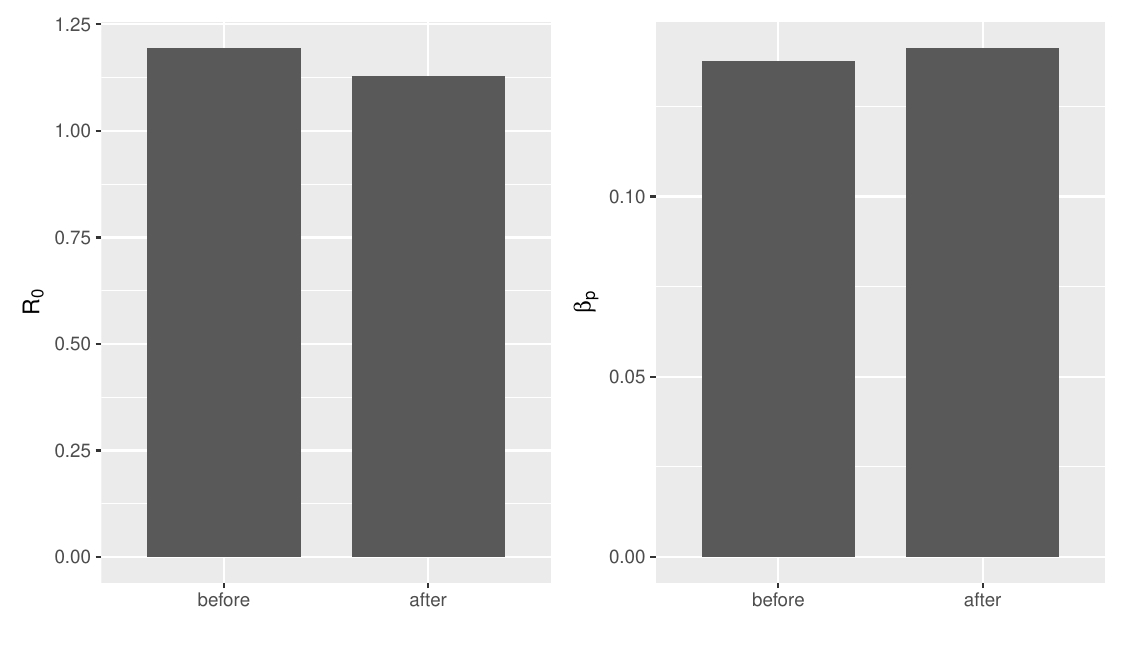}
\caption{The estimated parameter $\beta_p$ of the model and the basic reproduction number $R_0$ with $S_p=1$ analysing the data before and after the lockdown for Germany.}
\label{fig:BefAfterGermany}
\end{figure}

Now we want to determine the influence of governmental measures by estimating and comparing the parameters for the data before the governmental measures and afterwards. We have just faced 
the problem that there is only one maximum for the infectious population, therefore it is not possible with the method described in Section \ref{analyze} to determine the delay time $\tau_R$ for both time intervals. In order to get an estimate 
of the measures, we have used the time delay $\tau_R$, which we received for the analysis with the complete data set, for both time intervals, see Table \ref{tab:Europe}. In order to determine the influence of the lockdown, the data was divided into 
two parts, namely the one before and after the lockdown. 
Since the incubation period is approximately 5 days, a separation date from the lockdown date (23 March) plus 5 days (28 March) was used. In Figure \ref{fig:BefAfterGermany} we have plotted the parameter $\beta_p$ and the basic reproduction number 
$\protect\mathmbox{R_0 = \protect \beta_p\cdot S_p/(\gamma_R+\gamma_D)}$ before and after the lockdown for Germany. In order to be able to compare the basic reproduction number for the two cases, we have chosen $S_p$ equal to 1 for both cases.\\

No visible effect of the lockdown on the observed quantities is visible, which corresponds to the study by de Meunier \cite{Meunier2020} and which contradicts the study by Flaxman \cite{Flaxman2020}.

\section{Prediction}
\label{prediction}
It is of interest to predict things like number of maximal infectious, number of deceased and the infection fatality ratio as early as possible to help to make decisions about governmental measures. Our approach gives us 
the opportunity to make predictions of the maximal percentage of infectious $I_{p_{max}}$, the asymptotic percentage of dead $D_{p_{\infty}}$, the asymptotic percentage of recovered $R_{p_{\infty}}$. 
As mentioned in the previous section the infectious fatality ratio cannot be calculated. Therefore, we again will calculate the CFR which is given by $\protect\mathmbox{CFR= \protect D_{p_{\infty}}/(D_{p_{\infty}}+R_{p_{\infty}})}$.\\

Fortunately, it is possible to calculate $\lambda$ and $\delta$ without knowing the time delay $\tau_R$. Taking a look at the 
equations \ref{eq:SIRDdel2exact} it is obvious that $\delta$ is given by 
$-1/I_{p_x}$ and $\lambda$ by $-I_{p_{lnx}}/I_{p_x}$. Knowing $\lambda$, $S_{p_0}$ and $I_{p_0}$ we are able to calculate the asymptotic percentage of deaths $D_{p_{\infty}}$ respectively of recovered 
$R_{p_{\infty}}$ by putting in $S_{p_{\infty}}$ into the 
fitted functions for the deceased respectively recovered population against the susceptible population.\\

In order to get an estimate for each key date, we used the data up to the key date and prepared it as written in Section \ref{analyze}, whereby we re-determined the accessible population for each key date separately. 
After that, the desired quantities $D_{p_{\infty}}$ and $R_{p_{\infty}}$ were calculated as described above.\\

Afterwards we calculated for all dates a predictor for the corresponding quantity, for example  the absolute number of deceased $\tilde{D}_{\infty}=N_0\cdot D_{p_{\infty}}$ respectively recovered 
$\tilde{R}_{\infty}=N_0\cdot D_{p_{\infty}}$, the maximal number of 
infectious $\tilde{I}_{max}=I_{p_{max}} \cdot N_0$ or the case fatality ratio $\protect\mathmbox{\tilde{CFR}= \protect \beta_p\cdot S_p/(\gamma_R+\gamma_D)}$.\\

\subsection{Predictions using the data of the first wave}
\label{sec:predfirst}
\begin{figure}[h]
\centering
\hspace{-0.5cm}\includegraphics[width=6.5cm]{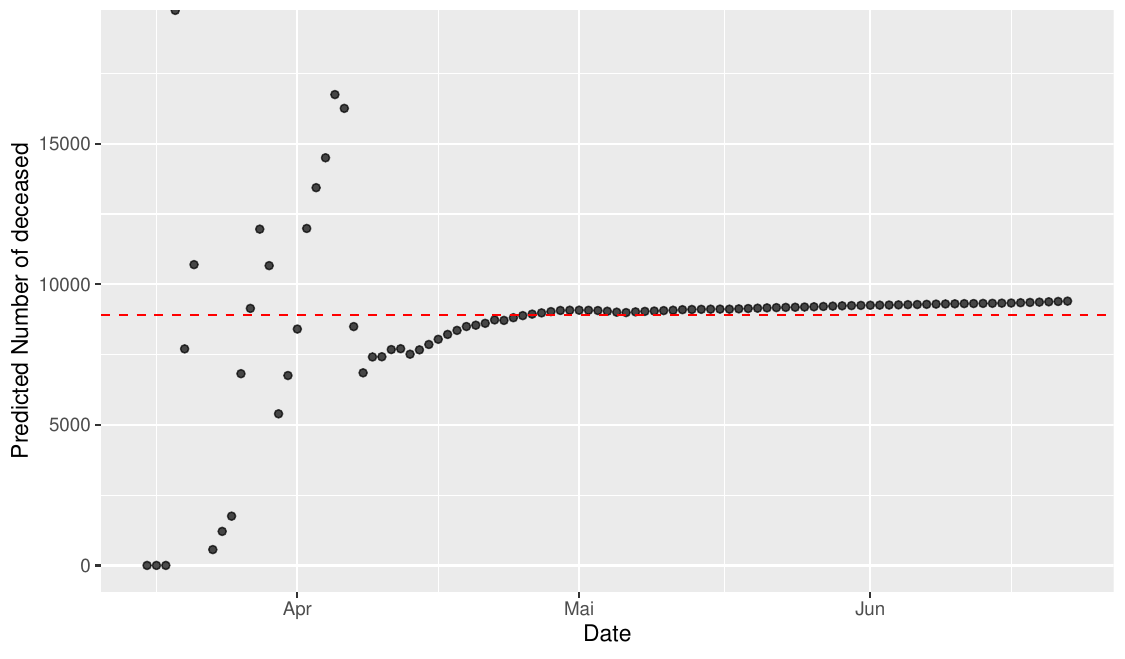}
\includegraphics[width=6.5cm]{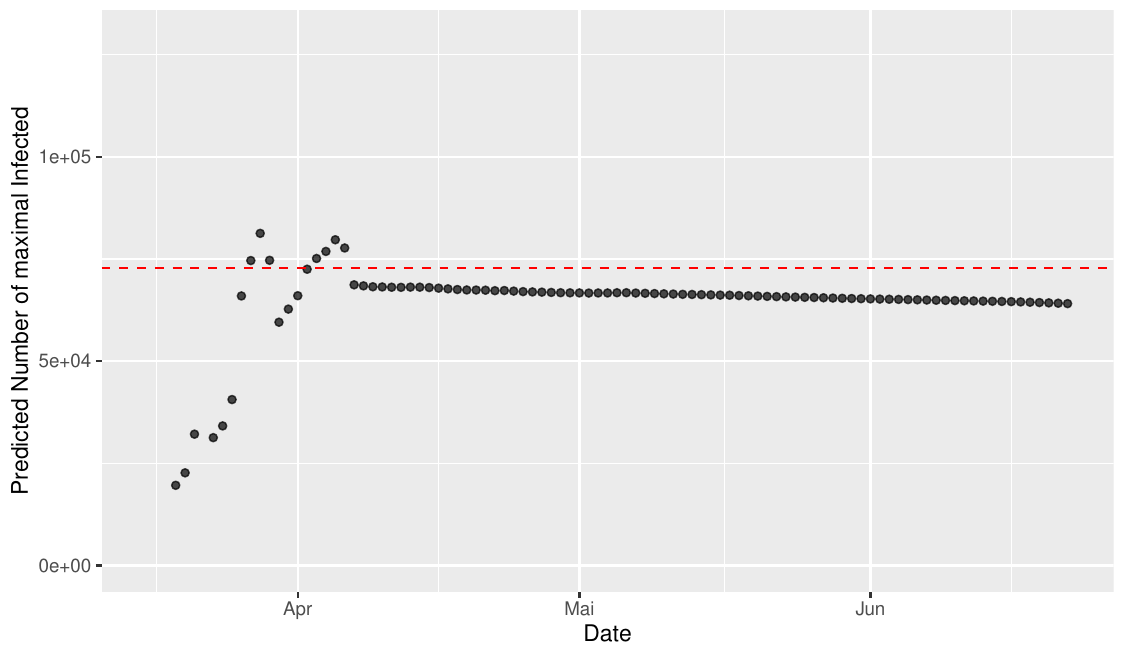}\\
\hspace{-0.5cm}\includegraphics[width=6.5cm]{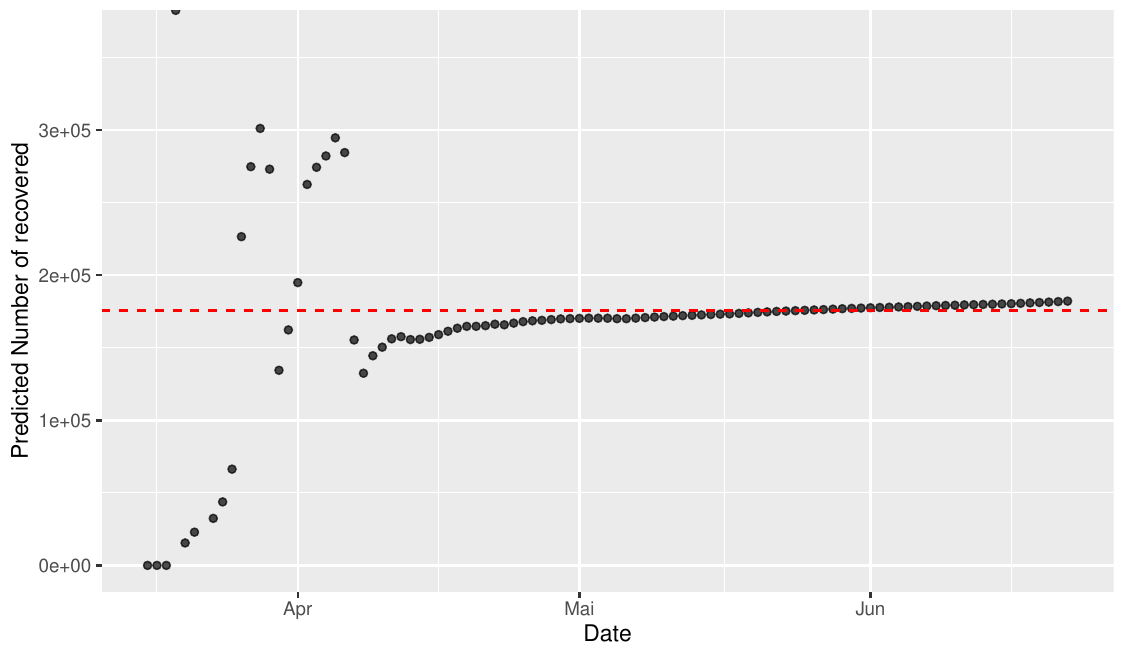}
\includegraphics[width=6.5cm]{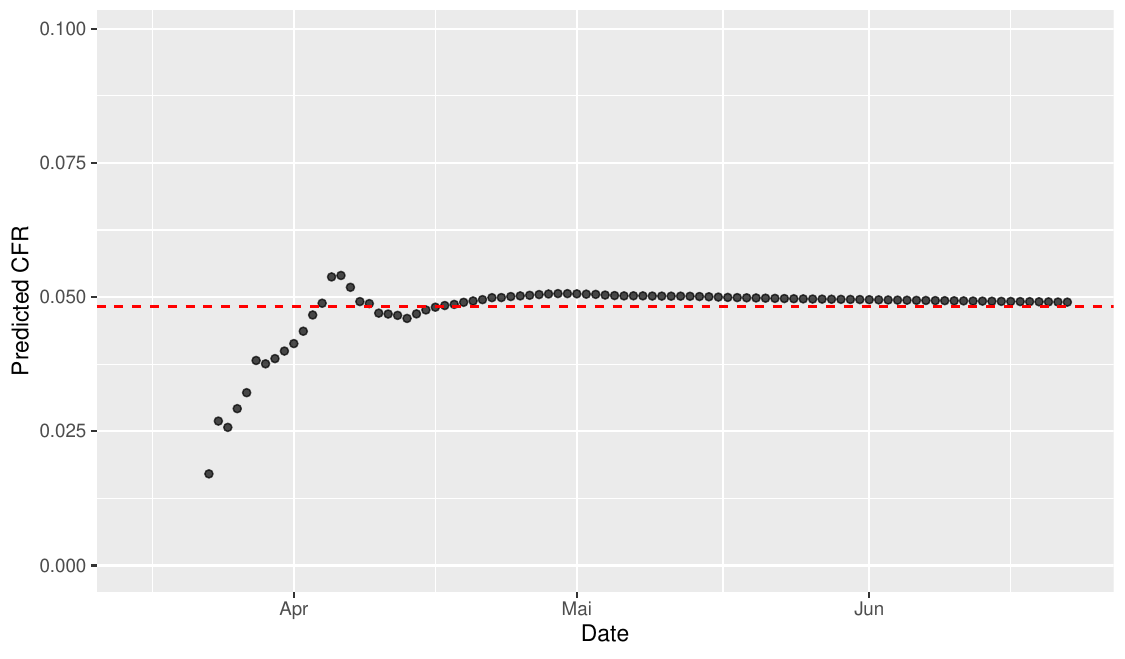}\\

\caption{The time dependence of the prediction of the number of deceased, of maximal infectious, of the recovered and the case fatality rate is shown. 
In every plot the dashed horizontal line indicates the corresponding value calculated from the data for the \protect $22^{nd}$ of June 2020.}
\label{fig:predGermanyd}
\end{figure}

We used the data until the $22^{nd}$ of June 2020  to analyse the first wave in Germany. In Figure \ref{fig:predGermanyd} we have plotted for every date with enough data for Germany the predictions of the number of deaths 
$\tilde{D}_{\infty}$, number of recovered $\tilde{R}_{\infty}$, the maximal number of infectious people $\tilde{I}_{max}$  and the CFR to get an impression if the predicted quantity is going to converge.\\

\begin{figure}[h]
\centering
\hspace{-0.5cm}\includegraphics[width=4.3cm]{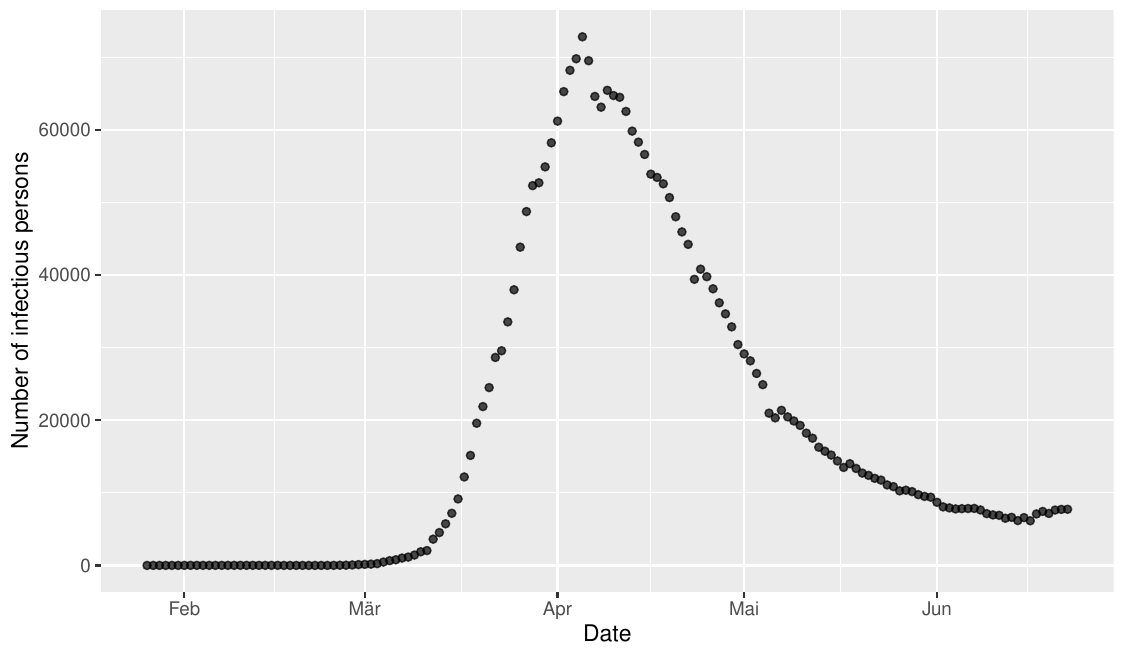}
\includegraphics[width=4.3cm]{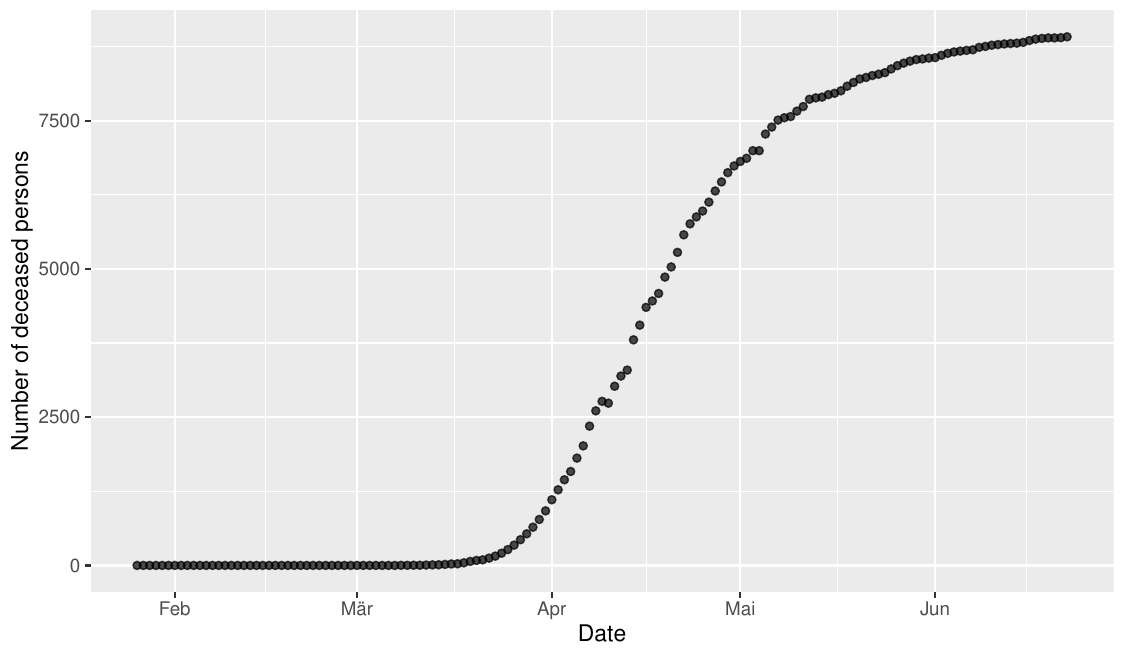}
\includegraphics[width=4.3cm]{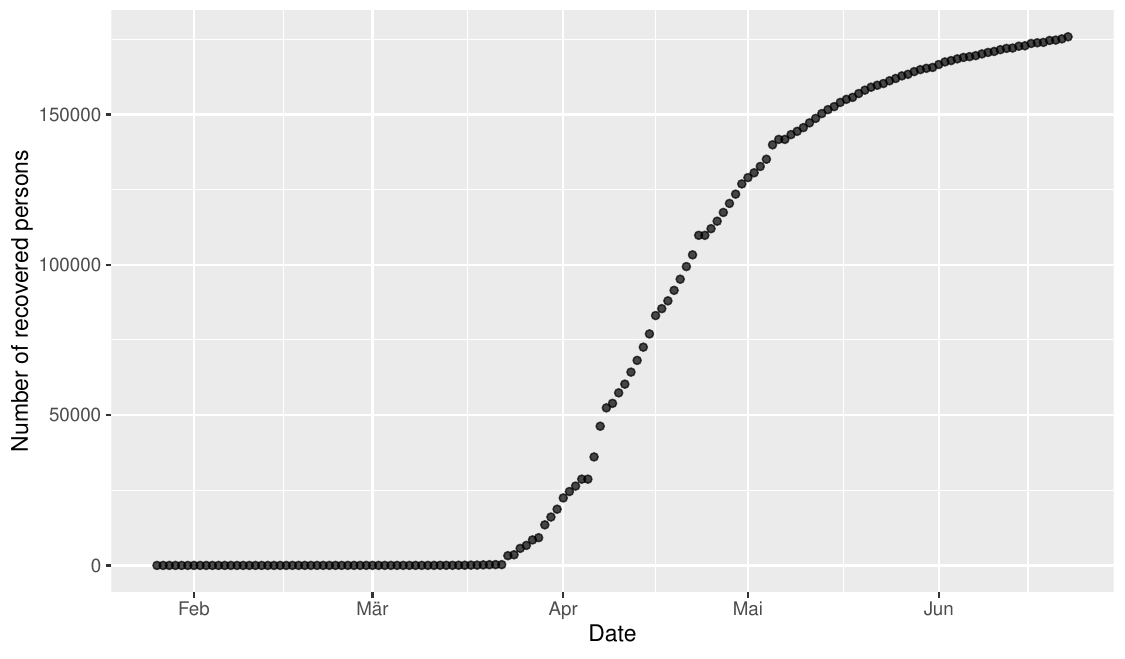}\\
\caption{From left to right we have plotted the data of Germany of the infectious $I$, the deceased $D$ and the recovered persons $R$ for the first wave. \label{fig:NumberGermany}}
\end{figure}

We see that in all plots in Figure \ref{fig:predGermanyd} that from around approximately the $10^{th}$ of April, the forecasts starts to converge against a value. By comparing the plots of the infectious 
population in Figure \ref{fig:NumberGermany} with the plots in Figure \ref{fig:predGermanyd} we see that a 
short period after the date of maximal infectious people the predictions start to converge. Thus, we are not able to make a valuable prediction of the maximal number of the infectious people. 
We have decided that we get a reasonable prediction if the following two conditions are met. The maximum number of infectious people must have already been exceeded and 4 consecutive predictions vary less than 10 percent. 
The first date by which this criterion is met we call first reasonable prediction date.\\

In Table \ref{tab:preda} we have summarized for the considered countries the actual number and the predicted number of deceased for the first reasonable prediction date and the last date of the data. The prediction of the 
number of deaths converges against  $9403$ which is a little bit higher than 
the actual number (date 22.06.2020) of deaths 8914 in Germany. As seen in Table \ref{tab:predb} the prediction for the CFR in April is closer to the actual value than the prediction of deaths in Germany.\\

\begin{table}[!htbp] \centering 
\scalebox{0.75}{
\makebox{
\begin{tabular}{@{\extracolsep{5pt}} lcccc} 
\\[-1.8ex]\hline 
\hline \\[-1.8ex] 
Country & Germany & Switzerland & Austria & Italy \\ 
\hline \\[-1.8ex] 
Actual date & 2020-06-22 & 2020-06-22 & 2020-06-22 & 2020-06-22 \\ 
actual Number of deaths & 8914 & 1956 & 693 & 34675 \\ 
number predicted at actual date & 9403 & 1987 & 711 & 36070 \\ 
Date of prediction & 2020-04-10 & 2020-04-04 & 2020-04-02 & 2020-05-04 \\ 
number of deaths at prediction date & 2736 & 715 & 168 & 29315 \\ 
number predicted at prediction date & 7424 & 1859 & 715 & 44013 \\ 
\hline \\[-1.8ex] 
\end{tabular}
}
}
  \caption{Comparison of the prediction with the actual value (June 22) of the number of deceased in Germany, Switzerland, Austria and Italy. We plotted the prediction at the actual date and 
  the prediction at the date where 4 predictions vary less than 10\% of the predicted value at this date and the maximal number of infectious people has exceeded.} 
  \label{tab:preda}
\end{table}

\begin{table}[!htbp] \centering 
\scalebox{0.75}{
\makebox{
\begin{tabular}{@{\extracolsep{5pt}} lcccc} 
\\[-1.8ex]\hline 
\hline \\[-1.8ex] 
Country & Germany & Switzerland & Austria & Italy \\ 
\hline \\[-1.8ex] 
Actual date & 2020-06-22 & 2020-06-22 & 2020-06-22 & 2020-06-22 \\ 
CFR at actual date & 0.048 & 0.063 & 0.041 & 0.158 \\ 
CFR predicted at actual date & 0.049 & 0.064 & 0.041 & 0.152 \\ 
Date of prediction & 2020-04-10 & 2020-04-06 & 2020-04-06 & 2020-03-25 \\ 
CFR at prediction date & 0.045 & 0.086 & 0.057 & 0.442 \\ 
CFR predicted at prediction date & 0.047 & 0.072 & 0.046 & 0.448 \\ 
\hline \\[-1.8ex] 
\end{tabular}
}
}
  \caption{Comparison of the prediction with the actual value (June 22) of the CFR in Germany, Switzerland, Austria and Italy.  The first reasonable prediction date was taken as described in Section \ref{sec:predfirst}.} 
  \label{tab:predb}

\end{table}

\subsection{Prediction of the number of deaths in the second and third wave}
\label{sec:2wave}
\begin{figure}[h]
\centering
\hspace{-0.5cm}\includegraphics[width=6.5cm]{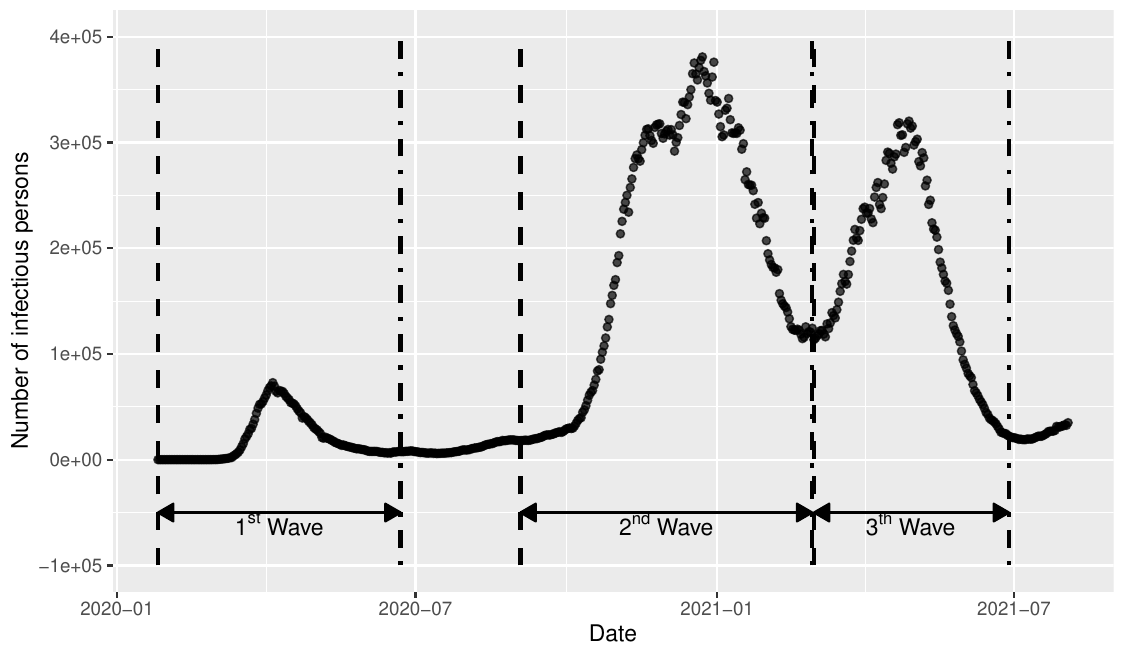}
\includegraphics[width=6.5cm]{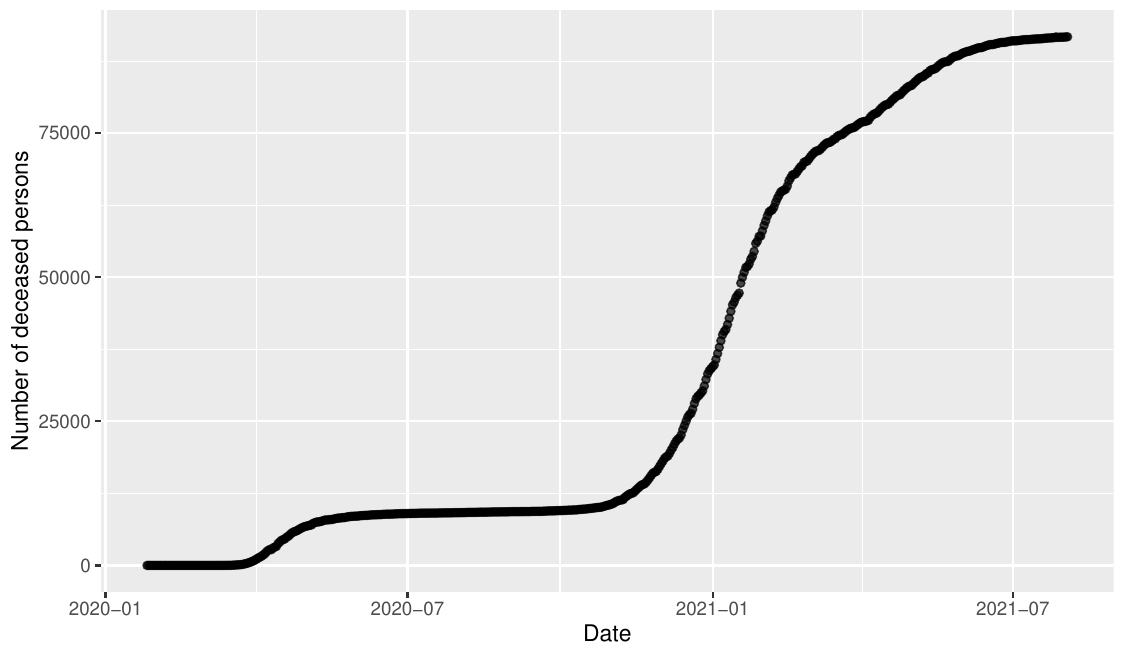}
\includegraphics[width=6.5cm]{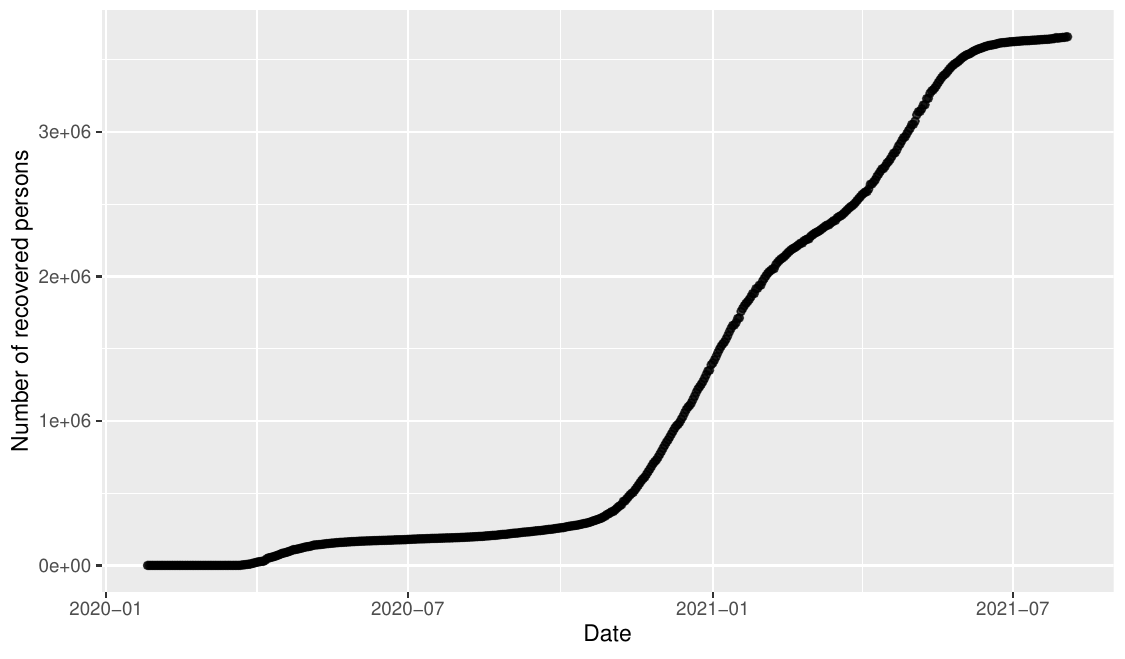}\\
\caption{From left to right we have plotted the data for Germany of the infectious $ I $, the deceased $ D $ and the recovered $ R $ population up to August 8, 2021. The start and end times of the first, second and third waves can be seen. These 
have been determined using the local minima in this plot.}
\label{fig:Plotalldata}
\end{figure}

\begin{figure}[h]
\centering
\hspace{-0.5cm}\includegraphics[width=6.5cm]{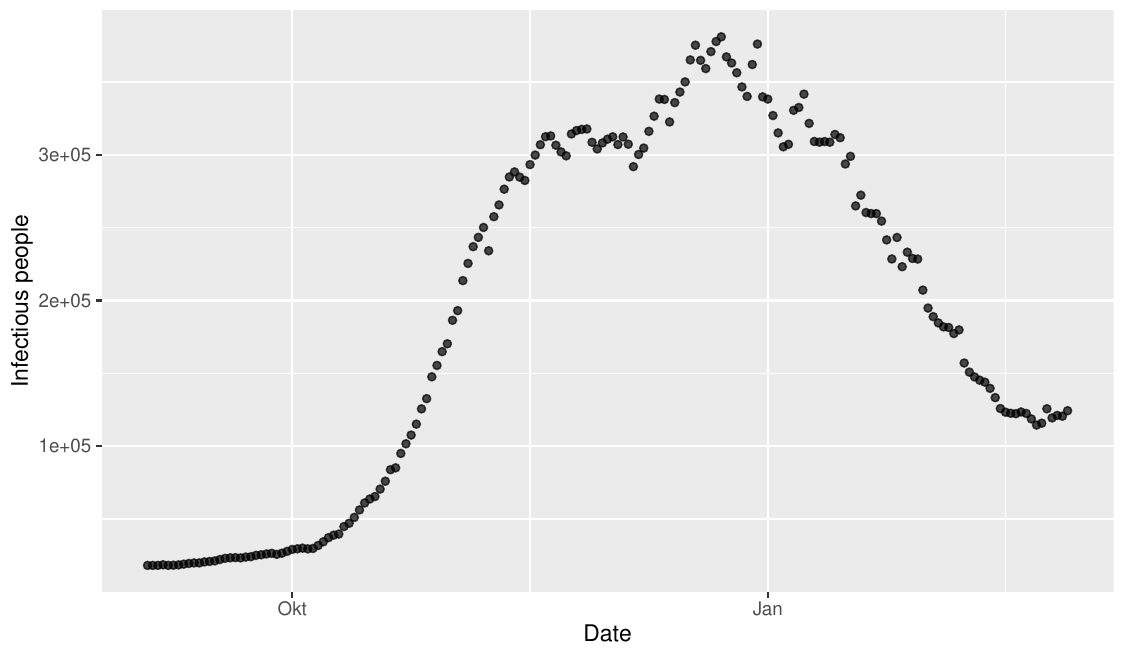}
\includegraphics[width=6.5cm]{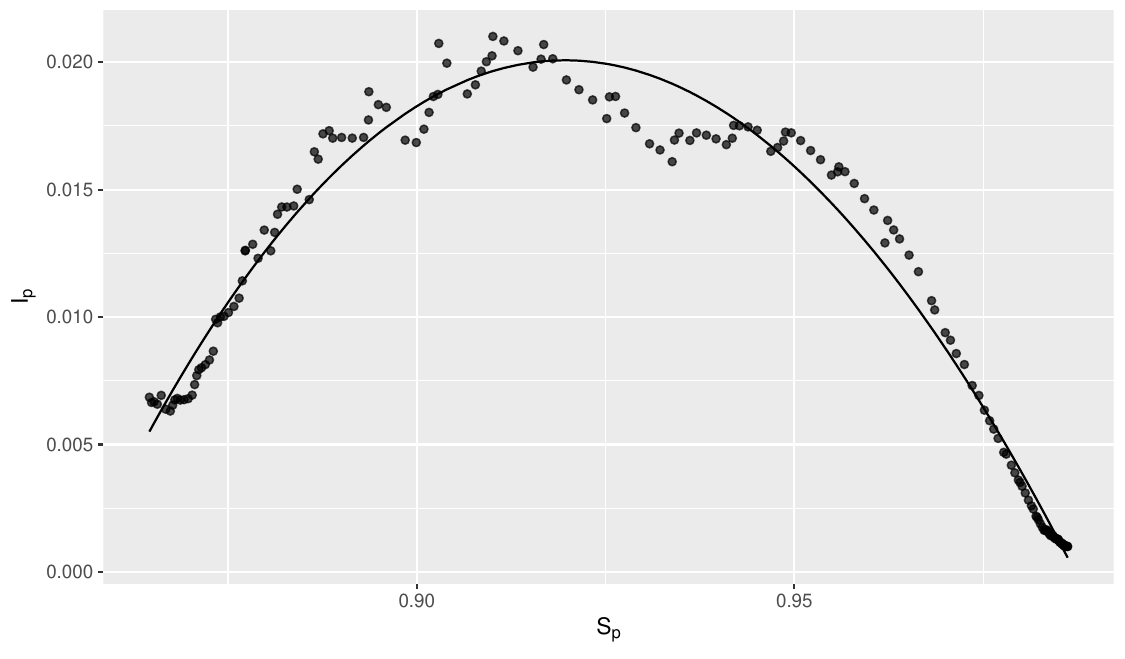}
\vspace{-10pt}
\\
\hspace{-0.5cm}\includegraphics[width=6.5cm]{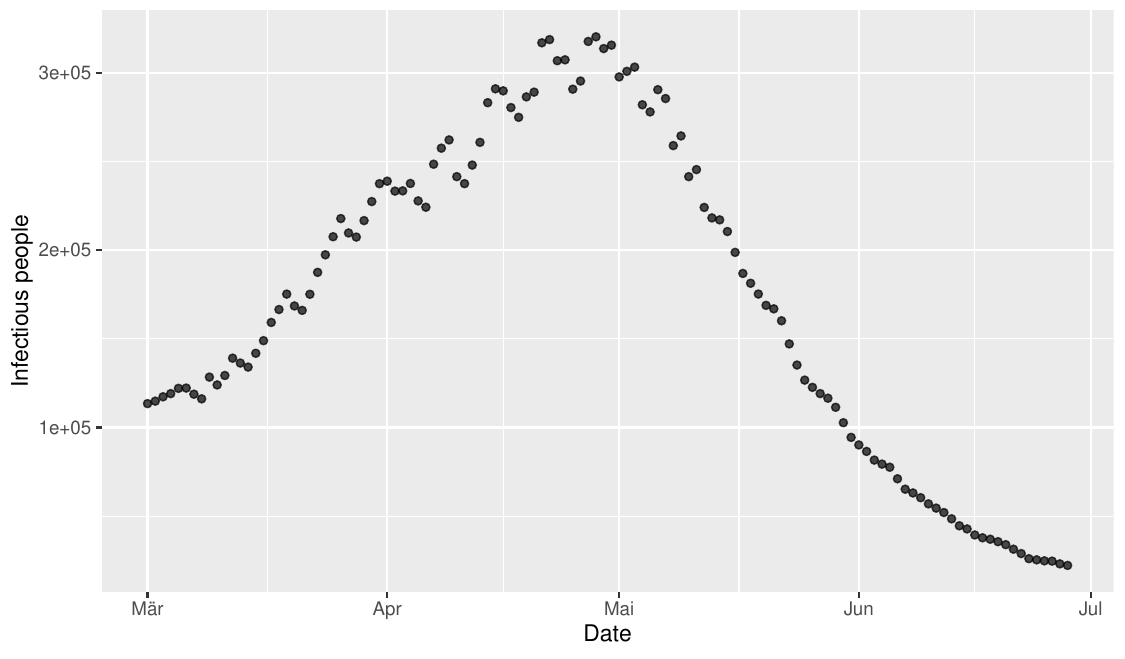}
\includegraphics[width=6.5cm]{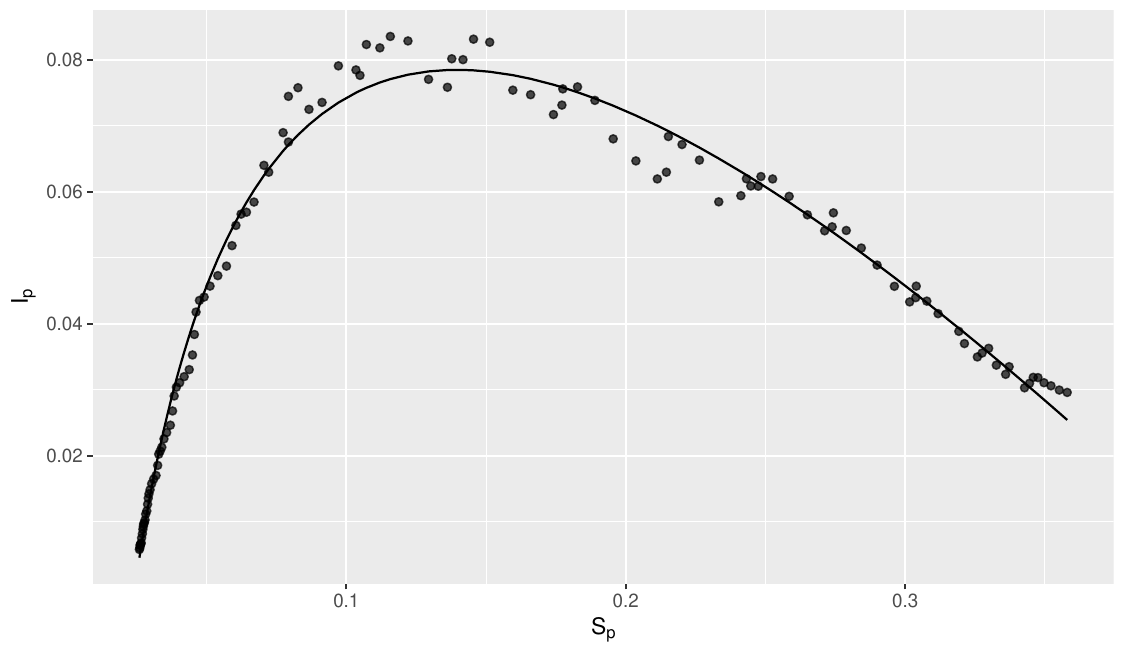}
\caption{In the left we have plotted the infectious  population against the time and on the right hand side the infectious  population  against the susceptible population. The upper part 
shows the two graphics for the second wave and the lower one for the 
third wave. It can be seen that the fitted functions are describing the data in the infectious versus the susceptible population plot quite well.}
\label{fig:wave2raw}
\end{figure}

We used the data until the $8^{th}$ of August 2021 to analyse the second and the third wave in Germany. In Figure \ref{fig:Plotalldata} we have plotted the time-dependence of 
the infectious $I$, the deceased $D$ and the recovered population $R$ for Germany. In the plot of the infectious, the 3 different waves are clearly visible. To separate the waves, 
we have calculated the date of local minimal values for the infectious population. Then we have chosen the appropriate minima to define the starting and ending date of the 3 different waves 
which we have summarized in the Table \ref{tab:wavedate} and are visible on the left-hand side of Figure \ref{fig:Plotalldata}. To get an impression of the quality of the fit, we 
have plotted in Figure \ref{fig:wave2raw} the infectious population versus time and versus the susceptible population and the fitted curve for the second and third wave. In the upper part of the figure, it can be seen that 
the infectious curve in the time representation of the second wave
has almost a double peak, the fit reproduces the curve quite well in the other representation and that the fit for the third wave also describes the curve well.\\

\begin{table}[!htbp] \centering
\scalebox{0.75}{
\makebox{
\begin{tabular}{@{\extracolsep{5pt}} lcc}
\\[-1.8ex]\hline 
\hline \\[-1.8ex] 
Wave & starting date & ending date\\
\hline \\[-1.8ex] 
1 & 16. January 2020 & 22. June 2020\\
2 & 2. September 2020 & 28. February 2021\\
3 & 29. February 2021 & 28. June 2021\\
\hline \\[-1.8ex] 
\end{tabular}
}}
\caption{The starting and ending date which we have received by looking at the mimima of the infectious people.\label{tab:wavedate}}
\end{table}

\begin{figure}[h]
\centering
\hspace{-0.5cm}\includegraphics[width=6.5cm]{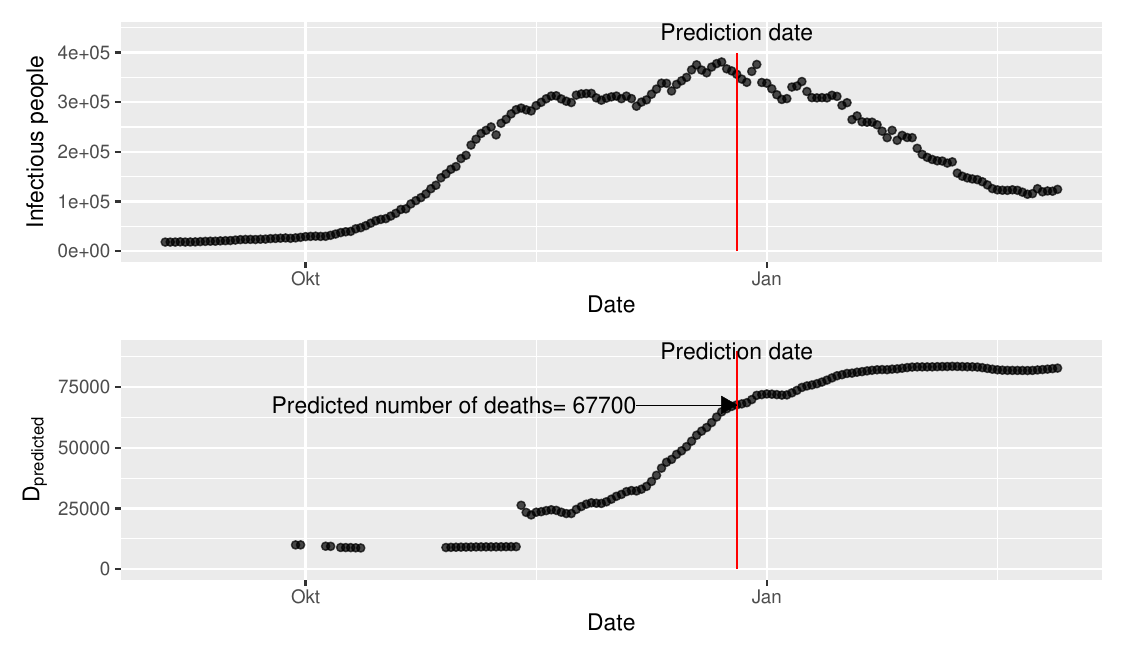}
\includegraphics[width=6.5cm]{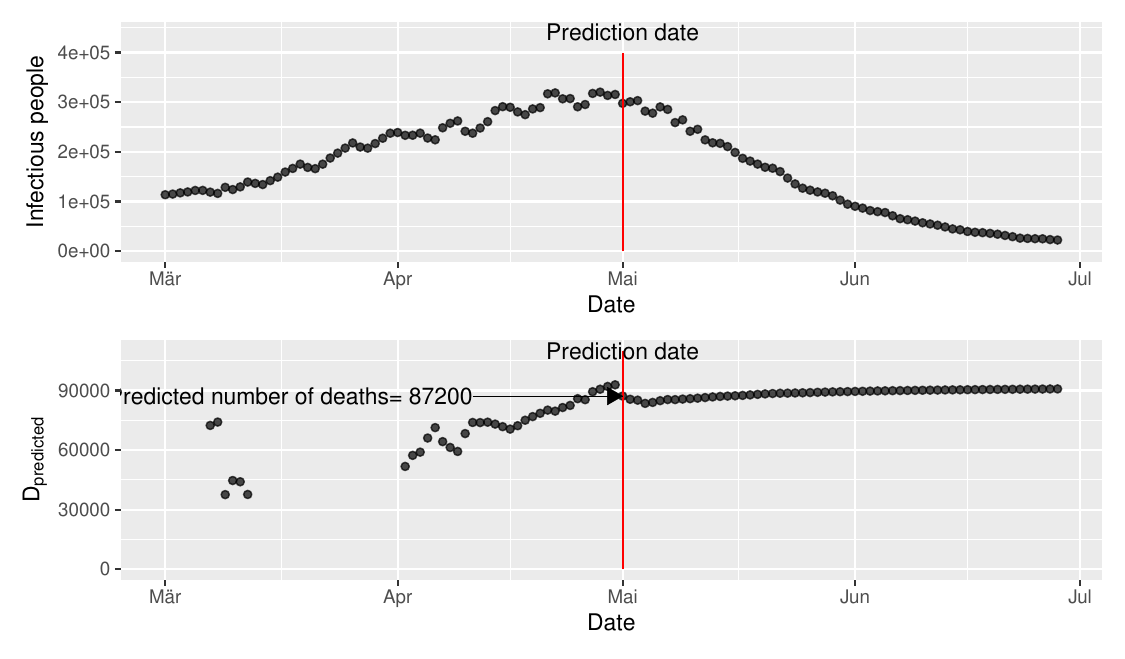}
\caption{In both graphs, the infectious against the time are shown in the upper part and the prediction is shown in the lower part. The left plots are from the data for the second 
wave, and the right part  shows the one for the third wave.}
\label{fig:waveallpred}
\end{figure}

To predict the number of deaths during the $2^{nd}$ resp. $3^{rd}$ wave we have used the same procedure as for the prediction for the $1^{st}$ wave. In Table \ref{tab:reswaves} the start times, the end times and the first reasonable forecast date with the corresponding number of deaths is shown. This table tells us that on the forecast date, about 30\% of the 
death toll of this wave died, and the forecast is about 20\% too low. In the Figure \ref{fig:waveallpred} it is visible that after the first reasonable prediction date the predicted number of deaths is still increasing. 

\begin{table}[!htbp] \centering
\scalebox{0.75}{
\makebox{
\begin{tabular}{@{\extracolsep{5pt}} cccc} 
\\[-1.8ex]\hline 
\hline \\[-1.8ex] 
Wave & 1 & 2 & 3 \\ 
\hline \\[-1.8ex] 
Starting date & 2020-01-26 & 2020-09-03 & 2021-03-01 \\ 
Number at the starting date & 0 & 9327 & 70926 \\ 
Date of prediction & 2020-04-10 & 2020-12-26 & 2021-05-01 \\ 
predicted number & 7424 & 67704 & 87193 \\ 
Number at the prediction date & 2736 & 30297 & 83292 \\ 
Ending date & 2020-06-22 & 2021-02-28 & 2021-06-28 \\ 
Number at the  ending date & 8914 & 70514 & 90883 \\ 
\hline \\[-1.8ex] 
\end{tabular} 
}}
  \caption{The approximate start and end date of all waves with the number of reported deaths at that moment are listed. Additionally the first reasonable date for a prediction with the predicted number of deaths is shown.} 
  \label{tab:reswaves} 
\end{table}

\begin{figure}[h]
\centering
\hspace{-0.5cm}\includegraphics[width=6.5cm]{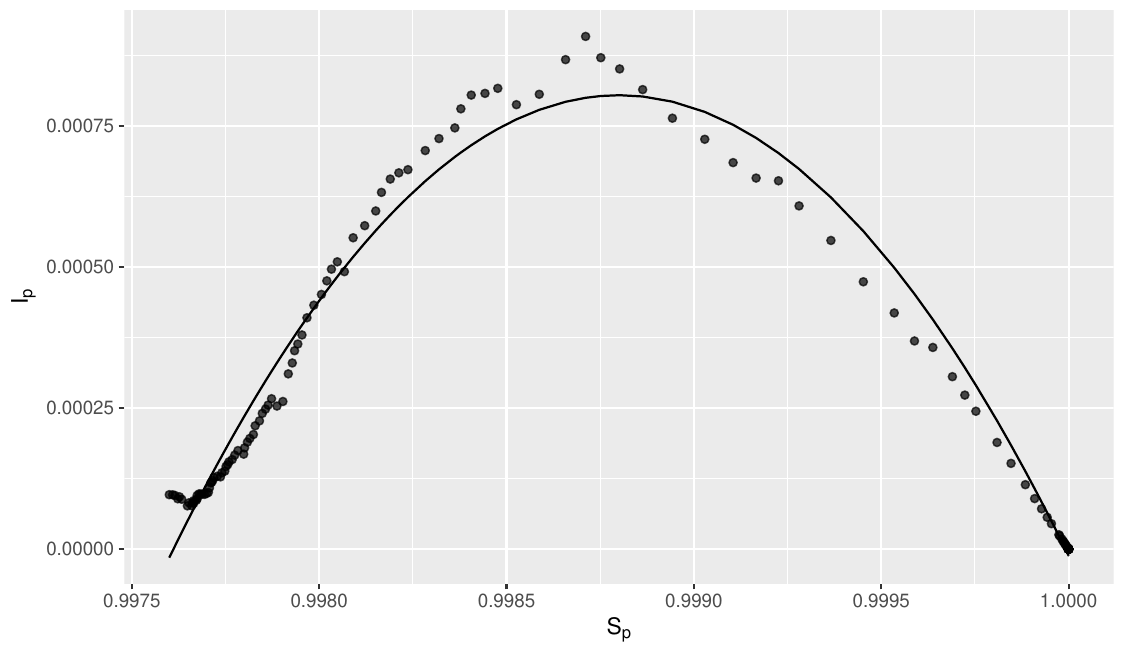}
\includegraphics[width=6.5cm]{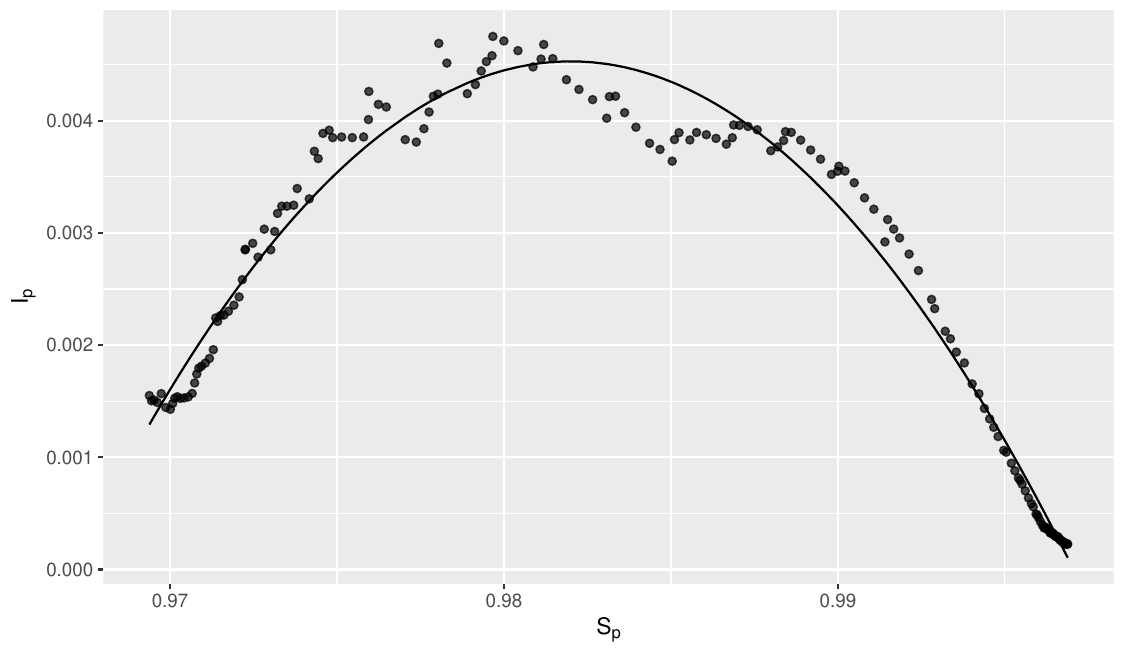}
\includegraphics[width=6.5cm]{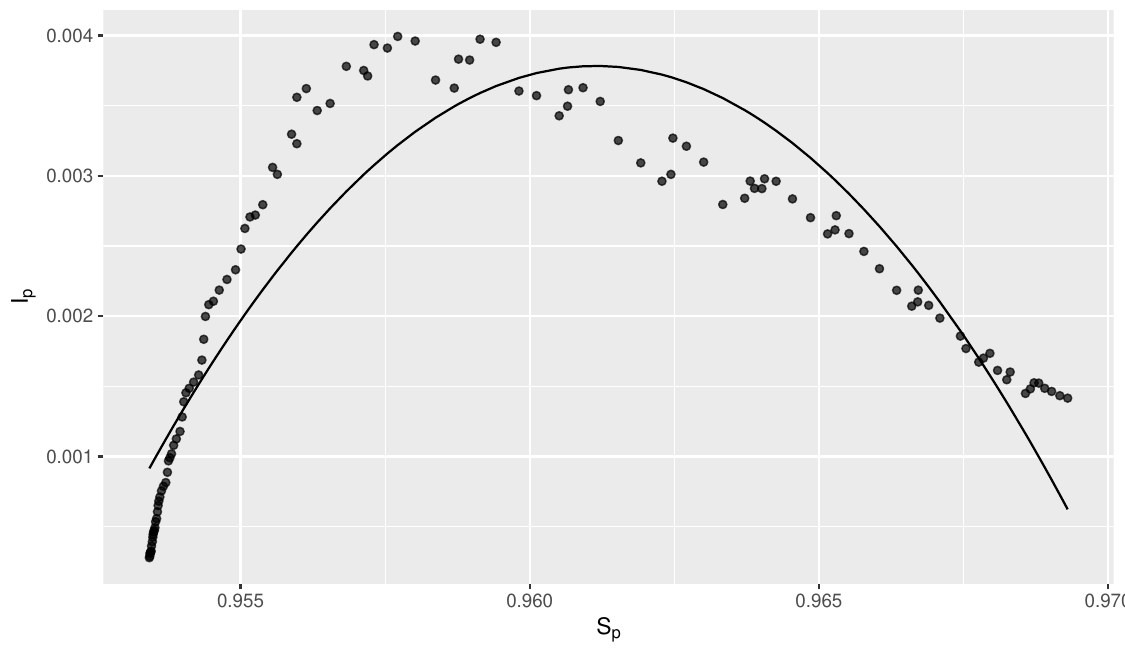}\\
\caption{From left to right, we have plotted the data and the fits of Germany for the first, the second and the third wave assuming that the whole population of Germany is susceptible. 
It is evident that the fits explains the first and second wave quite well in contrast to the third wave. This is maybe an effect of the vaccination.}
\label{fig:WavesNkonst}
\end{figure}

A possible effect of the vaccination may be seen taking a look at the plots of the infectious versus the susceptible population for the different waves. In Figure \ref{fig:WavesNkonst} we have plotted the fits by using the 
assumption that the whole population of Germany is susceptible. It is clearly evident that the fits for the first and second wave are a quite good approximation and that the fit 
for the third wave is terrible. Until the end 
of the second wave (28. February 2021) only approximately 5\% have received at least 1 dose of the vaccination, in contrast to the end of the third wave (28. June 2021) where approximately 50\% have received at least one dose 
of the vaccination.\\

\section{Discussion}

We proposed a SIRD-model including time-delay to describe data of the SARS-CoV-2 pandemic from different countries. The model has the remarkable property that we are able to find analytical results which allows us to 
determine the parameters of the model easily and to make different predictions. 
We have chosen to use the representations of infectious $I_p$, recovered $R_p$ and deceased population $D_p$ in dependence of the susceptible population $S_p$ instead of a representation in dependence of time. We used for all
analysis the data from Germany and we have seen that the data can be described quite well in this representation, but is not representing very well the time dependence. We were also able to show that a SEIRD respectively 
SIRD-model are not describing the data well.\\

Using the data for Germany up to June 22, 2020, the model estimates a time from detection of infection to death of 7.9 days, which agrees quite well with the value of different publications \cite{Italy,sousa2020} with a 
median from onset of symptoms to death, which ranges between 11 and 19 days. We also calculated that the time from onset of symptoms to recovery
to $8.1$ days, which is a good approximation compared to various studies \cite{healing20a,healing20b,healing20c}, which range between 10.6 and 21 days. \\

If we wanted to make a statement about governmental measures by comparing the parameter $\beta_p$ and the basic reproduction number $R_0$ of the model before and after the governmental measure, our results would 
indicate that the lockdown made no significant difference. This statement should be treated with caution, because we haven't calculated the error-bars.\\

We are able to make some predictions about the number of deaths and the case fatality ratio. We estimated for the first wave on April $10^{th}$, when 2736 persons 
in Germany died from COVID-19 that 7424 persons will die. The death toll on $22^{nd}$ of June 2020  was 8914, so we have underestimated the number of deaths on April $10^{th}$ by approximately 17\%. For the case fatality ratio we 
predicted 4.7\% on April $10^{th}$, compared to the 4.5\% data estimate at this date. Our estimate agrees very well with the data estimate on $22^{nd}$ of June 2020, which is 4.8\%. It is known that between 10\% to 90\% of the infected 
\cite{asymp20,asymp20a,Daym} are asymptomatic thus the infectious fatality ratio is approximately a factor 1.5 to 10 lower and is in accordance with the literature 
0.2\% \cite{Gangelt} respectively 1.6\% \cite{Erikstrup20,Ghisolfi2020}. It is surprising that even for the $2^{nd}$ and $3^{rd}$ wave in Germany the predictions are 
quite well and the data can be approximated very nicely by the obtained functions. By assuming that the whole population of Germany is susceptible our fits indicates that during the third wave, not the whole 
population of Germany is susceptible which may be an effect of the vaccination.\\

Additionally, we have investigated the data from Austria, Italy and Switzerland for the first wave. They can be analysed in the same way as we have analysed the data from Germany and the results 
except for Italy are not very different. All parameters excluding the parameter $\beta_p$ and the parameter $\tau_D$ are very similar.

\section*{Funding}

This research did not receive any specific grant from funding agencies in the public, commercial, or not-for-profit sectors.

\section*{Acknowledgment}

I would like to thank Luis Benet and Thomas Ott for their helpfull and stimulating comments.

\section*{Declaration of Competing Interest}

The author declare that he has no known competing financial interests or personal relationships that could have appeared to influence the work reported in this paper.

\bibliography{Sampleneu}

\end{document}